\documentclass[epjc3]{svjour3}
\journalname{Eur. Phys. J. C}
\usepackage[utf8]{inputenc}
\usepackage{booktabs}
\usepackage{amsmath}
\usepackage[separate-uncertainty=true, range-units = brackets]{siunitx}
\RequirePackage{graphicx}
\usepackage{url}
\usepackage{upgreek}
\usepackage{comment}
\usepackage{subcaption}
\usepackage{pdfpages}
\usepackage[title]{appendix}
\usepackage{lipsum}
\usepackage{setspace}
\usepackage{ulem}
\usepackage{lineno}
\usepackage[pdftex,bookmarks=true]{hyperref}
\newcommand{\jannis}[1]{{\color{black}{#1}}}
\newcommand{\jannisfourty}[1]{{\color{black}{#1}}}

\newcommand{\AL}[1]{{\color{black}{#1}}}
\newcommand{\AS}[1]{{\color{black}{#1}}}%
\newcommand{\ChWe}[1]{{\color{black}{#1}}}%

\newcommand{\tritium}{\mathrm{T}}
\newcommand{\hydrogen}{\mathrm{H}}
\newcommand{\deuterium}{\mathrm{D}}

\newcommand{\TT}{{\tritium_2}}
\newcommand{\HH}{{\hydrogen_2}}
\newcommand{\HT}{{\mathrm{HT}}}
\newcommand{\DT}{{\mathrm{DT}}}
\newcommand{\HeT}{{^3\mathrm{HeT}^+}}
\newcommand{\HeH}{{^3\mathrm{HeH}^+}}
\newcommand{\HeD}{{^3\mathrm{HeD}^+}}

\begin{document}

\title{Improved treatment of the T$_2$ molecular final-states uncertainties for the KATRIN neutrino-mass measurement}

\onehalfspacing
\titlerunning{\parbox{9cm}{Improved treatment of the T$_2$ molecular final-states uncertainties}} 
\authorrunning{S.~Schneidewind, \dots, A.~Saenz}
%authors
\institute{%
Institute for Nuclear Physics, University of M\"{u}nster, Wilhelm-Klemm-Str. 9, 48149 M\"{u}nster, Germany\label{a}
\and Institut f\"{u}r Physik, Humboldt-Universit\"{a}t zu Berlin, Newtonstr. 15, 12489 Berlin, Germany\label{b}
\and Institute of Experimental Particle Physics~(ETP), Karlsruhe Institute of Technology~(KIT), Hermann-von-Helmholtz-Platz 1, 76344 Eggenstein-Leopoldshafen, Germany\label{c}
}
\thankstext{email1}{e-mail: sonja.schneidewind@uni-muenster.de}
\thankstext{email2}{e-mail: alejandro.saenz@physik.hu-berlin.de}
% Authors:
\author{%
S.~Schneidewind\thanksref{a,email1}
\and J.~Sch\"{u}rmann\thanksref{b,a}
\and A.~Lokhov\thanksref{c,a}
\and A.~Saenz\thanksref{b,email2}
\and C.~Weinheimer\thanksref{a}
}

\date{\today}
\maketitle

\begin{abstract}
The KArlsruhe TRItium Neutrino experiment (KATRIN) aims to determine the effective mass of the electron antineutrino via a high-precision measurement of the tritium $\upbeta$-decay spectrum in its end-point region.
The target neutrino-mass sensitivity of $\SI{0.2}{\electronvolt\per c^2}$ at 90~\% C.L. can only be achieved in the case of high statistics and a good control of the systematic uncertainties.
One key systematic effect originates from the calculation of the molecular final states of T$_2$ $\upbeta$ decay.
In the first neutrino-mass analyses of KATRIN the contribution of the uncertainty of the molecular final-states distribution (FSD) was estimated via a conservative phenomenological approach to be $\SI{2e-2}{\square\electronvolt\per c^4}$.
In this paper a new procedure is presented for estimating the FSD-related uncertainties by considering the details of the final-states calculation, {\it i.\,e.\ }the uncertainties of constants, parameters, and functions used in the calculation as well as its convergence itself as a function of the basis-set size used in expanding the molecular wave functions.   
The calculated uncertainties are directly propagated into the experimental observable, the squared neutrino mass $m_\nu^2$. With the new procedure the FSD-related uncertainty is constrained to $\SI{1.3e-3}{\square\electronvolt\per c^4}$, for the experimental conditions of the first KATRIN measurement campaign.

\end{abstract}
\section{Introduction}

The Karlsruhe Tritium Neutrino experiment (KATRIN) \cite{DesignReport}\cite{KATRINHardwarePaper} aims at determining the neutrino mass $m_\nu$ by precisely measuring the integrated electron-energy spectrum of the superallowed molecular tritium $\upbeta$ decay near the spectrum's end point at about \SI{18.57}{\kilo\electronvolt}. KATRIN combines an ultra-luminous Windowless Gaseous Tritium Source (WGTS) providing a $\upbeta$-decay rate of up to \SI{1e11}{Bq} \cite{Babutzka_2012} with a large spectrometer of MAC-E-filter \cite{mac-e} type transmitting electrons above an adjustable energy threshold with ${\cal O}(1)\,\mathrm{eV}$ width. The target sensitivity is $m_\nu <\SI{0.2}{\electronvolt / c^{2}}$ at $90\%$ C.L. after five years of data taking. The first four-week science campaign during spring 2019 (KNM1) yielded a limit of $\SI{1.1}{\electronvolt / c^2}$ \cite{KNM1Paper}, and the first two science campaigns (KNM1 and KNM2) set an upper limit of  $m_\nu < \SI{0.8}{\electronvolt / c^{2}}$ ($90\%$ C.L.) \cite{NaturePaper}.

Although the uncertainty of the neutrino mass extracted from these first measurement campaigns is dominated by the statistical error, a good control of the systematic effects and related uncertainties will be required in the future for reaching the target sensitivity of KATRIN. The main experimental uncertainties are connected to the distortions of the shape of the spectrum by various background effects, electrons' starting potential, and scattering in the source, as well as the transmission properties of the spectrometer. Another source of uncertainty stems from the final states of molecular tritium (T$_2$) $\upbeta$ decay. This final-states distribution (FSD) enters the computation of the differential $\upbeta$-decay spectrum that is used in the fit from which the squared neutrino mass is extracted. An FSD that is sufficiently accurate for the analysis of neutrino-mass experiments is so far only available from {\it ab initio} calculations \cite{PhysRevLett.84.242}. As was shown in \cite{doi:10.1146/annurev.ns.38.120188.001153}, a not considered variance $\sigma^2$ of the energy scale of the measured $\beta$ spectrum leads approximately to a shift in the squared neutrino mass extracted from the experiment according to $\Updelta m_\nu^2 = -2\cdot \sigma^2$. From this relation a first estimate of the FSD-related uncertainty on the neutrino mass can be derived: an uncertainty of the FSD used in the data analysis being described by the difference of the true variance of the FSD $\sigma^2_\mathrm{true}$ and of the FSD variance $\sigma^2_\mathrm{fit}$ used in the data analysis leads to an additional contribution to the neutrino-mass uncertainty of $|\Updelta m^2_\nu |= 2 \cdot |\sigma^2_\mathrm{true}-\sigma^2_\mathrm{fit}|$.

In this work the systematic uncertainties due to the molecular final states of T$_2$ ($\DT$, $\HT$) $\upbeta$ decay are revisited. In contrast to the previous uncertainty estimation that was based on a fully phenomenological approach \cite{AnalysisPaper}, the new procedure involves a detailed investigation of the various sources of uncertainties which enter the molecular FSD calculation. 
This includes the uncertainties from the use of a finite basis set in the {\it ab initio} calculation, from adopted approximations like the sudden approximation, and uncertainties on fundamental constants.
Different FSDs are generated, {\it e.\,g.}, by a systematic increase of the basis set or the inclusion (omission) of corrections to the adopted approximations. The comparison of the resulting squared neutrino masses $m^2_\nu$ -- the KATRIN observable -- that are obtained by a fit to a reference $\upbeta$ spectrum yields an effective shift $\Delta m_\nu^2$.
\footnote{The fact that an electron neutrino $\nu_e$ is a mixture of the different mass eigenstates $m(\nu_i)$ described by the neutrino mixing matrix elements $|U^2_{ei}|$ is neglected here, an ``effective electron antineutrino mass''  $m_\nu^2 := m^2(\nu_e) := \sum_i |U^2_{ei}| \cdot m^2(\nu_i)$ is used instead.} 
The different $\Delta m_\nu^2$ contributions are then added to a total systematic uncertainty of $m_\nu^2$ due to the FSD.

The FSD-related systematic uncertainty is determined in this study for the measurement conditions of KATRIN's first science campaign KNM1, since the adopted reference $\upbeta$ spectrum is generated with the corresponding experimental parameters like the source temperature or isotopologue distribution. 
A direct re-evaluation of the uncertainty using this new approach is, however, not fully possible in the case of the FSD that was used for the analysis of the first two KATRIN science campaigns (named KNM1 FSD). Since the KNM1 FSD was constructed by adopting the best input data available in literature at that time, there is, {\it e.\,g.}, no common basis set used for different states and the adopted corrections partially stem from calculations using again different basis sets \cite{valerian_paper}. Furthermore, a full reconstruction  of the KNM1 FSD (in the sense of a re-evaluation of all values from scratch) is not possible, since some of the basis-set parameters were unavailable in literature. Therefore, a fully consistent convergence study for the KNM1 FSD itself, as is required for the here proposed new uncertainty analysis, is not possible. It should be emphasised that the KNM1 FSD itself is very accurate, but the uncertainty determination of it is not easily possible. The re-assessment of the uncertainty is thus performed based on a newly generated pseudo-KNM1 FSD (introduced in detail in sec. \ref{sec:5.3}), adapted to the new uncertainty-analysis procedure. The procedure established in this work will be repeated for the analysis of future KATRIN measurement campaigns individually, i.\,e.\ depending on the experimental conditions of each campaign. Furthermore, an improved new FSD (not presented in this paper) will be adopted in the analysis of future campaigns that avoids the need for a pseudo FSD in the uncertainty estimate. 

This paper is structured as follows: 
the model of the tritium $\upbeta$-decay spectrum of KATRIN and the analysis procedure for the experimental conditions of the first measurement campaign are described in sec.\,\ref{sec:model}.
The impact of the molecular final states on the $\upbeta$-decay spectrum and the general procedure of the FSD computation are given in sec.\,\ref{sec:FSD_theory}. In sec.\,\ref{sec:quantifying} the uncertainties of the FSD computation are introduced and the previous approach to the uncertainty estimation is summarised. The new procedure to systematically assess the FSD uncertainty is explained in sec.\,\ref{sec:new_uncertainty}.
The resulting systematic contributions of the FSD calculation to the uncertainty of $m_\nu^2$ are presented in sec.\,\ref{sec:results}. A brief summary and outlook is given in sec.\,\ref{sec:summary}. Details on the choice of parameters for the FSD computation in the present work can be found in app. \ref{app:FSD_parameters}. For a list of all FSDs mentioned in this paper, see app. \ref{glossary}.

%----------------------------------------------------------------------------------------------------
\section{Model of the experimental $\upbeta$-decay spectrum\label{sec:model}}

In tritium neutrino-mass measurements the squared neutrino-mass parameter $m_\nu^2$ is inferred by fitting a model spectrum to a measured kinetic energy spectrum of the electrons emitted during $\upbeta$ decay.
In KATRIN all electrons with a kinetic energy above a specific threshold are detected. This threshold energy is scanned to obtain an integrated spectrum of $\upbeta$ electrons. The model of the integrated spectrum measured by KATRIN is described by a convolution of the theoretical differential $\upbeta$-decay spectrum $R_\upbeta (E)$, given by Fermi's Golden Rule, with the experimental response function $f_\text{calc}(E, qU)$~\cite{Kleesiek:2018mel},

\begin{equation}
    R_\text{calc}(qU)=A_\text{s}\cdot N_\text{T}\cdot \int_{qU}^{E_\text{0}}{R_\upbeta (E) \cdot f_\text{calc}(E, qU)\:dE} \:+\: R_\text{bg}.
    \label{general-spectrum}
\end{equation}

Here, $E$ is the kinetic energy of the $\upbeta$ electron and $qU$ is the retarding-voltage set point, which defines the energy threshold for the electrons transmitted by the spectrometer, $q=-e$ is the charge of the electron. $R_\text{bg}$ stands for a constant background rate which is a free parameter in the fit. The response function, $f_\text{calc}(E, qU)$, takes into account the energy losses due to scattering and synchrotron radiation, as well as the spectrometer transmission properties based on the magnetic fields along the beamline. $N_\text{T}$ is the number of tritium atoms $N_\text{T, abs}$ multiplied by the solid acceptance angle and the detector efficiency. $N_\text{T, abs}$ is defined via $N_\text{T, abs}=2\cdot\epsilon_\text{T}\cdot \rho d \cdot A$, with $A$ being the cross-section area of the flux tube within the windowless gaseous tritium source (WGTS), and $\epsilon_\text{T}=[N_{{\text T}_2}+\frac{1}{2}(N_\text{HT}+N_\text{DT})]/\sum_i N_i$ the tritium purity. $N_i$ is the number of molecules of one of the isotopologues T$_\text{2}$, DT, D$_\text{2}$, HT, HD and H$_\text{2}$. The amount of HT and DT in the source is described by the HT/DT ratio $\upkappa = N_\text{HT}/{N_\text{DT}}$ \cite{LARA}. At the WGTS, a constant tritium flow is achieved by continuously injecting molecular tritium gas of high purity in the midpoint of the beam tube. It is then diffusing to both sides, where it is pumped out. The column density $\rho d$ of the source is the integrated tritium density along the length $d=\SI{10}{\meter}$ of the source cryostat. $N_\text{T}$ can vary between different measurement campaigns. Finally, $A_\text{s}$ in eq.\,\ref{general-spectrum} is the relative signal amplitude, it is a free parameter in the fit. The other two fit parameters, the end point $E_\text{0}$ and the squared neutrino mass $m_\nu^2$, enter eq.~\ref{general-spectrum} via $R_\upbeta (E)$ (see sec.~\ref{sec:betaspectrum}, eq.~\ref{eq:introduction:beta_decay_fermi}).

For the FSD-uncertainty studies presented in this work, the integrated spectrum $R_\text{calc}(qU)$ is evaluated at discrete retarding-voltage set points $qU$. In sec.~\ref{run-condition}, the experimental parameters as well as the features of the spectrum model used in the analysis are described. The parameters correspond to the first KATRIN science campaign (KNM1). In sec.~\ref{generation}, a description of the used reference Asimov Monte Carlo data set is given. Finally, information on the fit methods is given in sec.\,\ref{fit}.

\subsection{Experimental conditions of the first science campaign (KNM1)\label{run-condition}}
\noindent\textit{\textbf{Tritium source parameters}}
The key source parameters of KNM1 which are used for the present FSD-uncertainty studies are listed in tab. \ref{knm1-source}, more details can be found in \cite{KNM1Paper}\footnote{There are small deviations between the values stated in \cite{KNM1Paper} and the values used here, because the latter ones are based on the most recent knowledge on the parameters during KNM1.}\\
\begin{table}
\centering
\caption{Source parameters during KNM1 as used in the present analysis. \label{knm1-source}}
\begin{tabular}{llll}
\toprule
Parameter& Value & Unit  \\
\midrule
 Column density $\rho d$   &  $1.11\times 10^{17}$ & $\text{mol}\cdot \text{cm}^{-2}$\\
Temperature $T$ & $30.1$ & K \\
Tritium purity $\epsilon_\text{T}$ & $97.6$ & $\%$\\
HT/DT ratio $\upkappa$ & 3.329&-\\
\bottomrule
\end{tabular}
\end{table}

\noindent\textit{\textbf{Spectrometer and beamline conditions}}
The $\upbeta$-decay electrons are emitted in a high magnetic field $B_\text{src}$ in the source. The magnetic field guides them adiabatically towards the spectrometers where they are filtered via the retarding potential energy $qU$. The filter width $\Delta E /E$ is defined by the ratio of the minimum magnetic field $B_\text{min}$ in the spectrometer's analysing plane and the maximal magnetic field in the beamline, $B_\text{max}$. In the configuration of KNM1~\cite{KNM1Paper} the fields have the values listed in tab. \ref{knm1-beamline}, leading to $\Delta E = \SI{2.8}{\electronvolt}$ at the $\upbeta$-decay end point $E_\text{0}$. The values from tab. \ref{knm1-beamline} are used as input values for generating the Monte-Carlo data in the present studies.

\begin{table}
\centering
    \caption{Magnetic fields along the beamline during KNM1, used as input for the Monte-Carlo data entering the FSD-convergence studies. \label{knm1-beamline}}
\begin{tabular}{llr}
\toprule
Parameter& Value & Unit  \\
\midrule
Field in WGTS $B_\text{src}$ & 2.51 & T\\
Maximum field along beamline $B_\text{max}$ & 4.24 & T\\
Minimum field $B_\text{min}$ & 0.63 & mT\\
\bottomrule
\end{tabular}
\end{table}

\noindent\textbf{\textit{Spectrum model and systematic corrections}}
For the description of the tritium $\upbeta$-decay spectrum in this study a fully relativistic Fermi function is used. Radiative corrections due to virtual and real photons, described in more detail in \cite{PhysRevC.28.2433} and \cite{Kleesiek:2018mel}, are applied. Furthermore, synchrotron energy losses \cite{Groh2015_1000046546} of the electrons in the high magnetic fields are taken into account, while Doppler broadening is not applied\footnote{The Doppler broadening only introduces a small Gaussian broadening of the spectrum which is on the order of KATRIN's finite energy resolution. It was tested in several places that the inclusion of the Doppler broadening does not have significant impact on the fit results.}.\\
The experimental response function is influenced by energy losses of the electrons in the source due to scattering. An energy-independent inelastic cross-section of $\SI{3.64e-18}{\square\centi\meter}$ \cite{KNM1Paper} at $E_\text{0}$ is assumed. In the simulation, electrons with up to seven scatterings are taken into account. The dependence of the scattering probability of the electron pitch angle with regard to the magnetic field lines due to the increase of the mean path length with increasing pitch angle \cite{Kleesiek:2018mel} is neglected in the study, the angle-averaged values are used instead.

\subsection{Generation of Monte-Carlo data\label{generation}}
    \begin{table}
    \centering
\caption{Input parameters for the KNM1 Monte-Carlo data. $A_\text{s}$, $E_\text{0}$, $R_\text{bg}$ and $m_\nu^2$ are the four free fit parameters in all fits performed for the FSD-uncertainty estimation.\label{fit-parameter}}
\begin{tabular}{llr}
\toprule
Parameter& Value & Unit  \\
\midrule
Relative signal amplitude $A_\text{s}$ &  1.0 & -\\
End point $E_\text{0}$ & 18573.7 & eV\\
Background rate $R_\text{bg}$ & 2.5 & mcps/pixel\\
Squared neutrino mass $m_\nu^2$ & 0 & eV$^2$\\
\bottomrule
\end{tabular}
\end{table}

\begin{figure}
      \centering
      \includegraphics[width=0.6\textwidth]{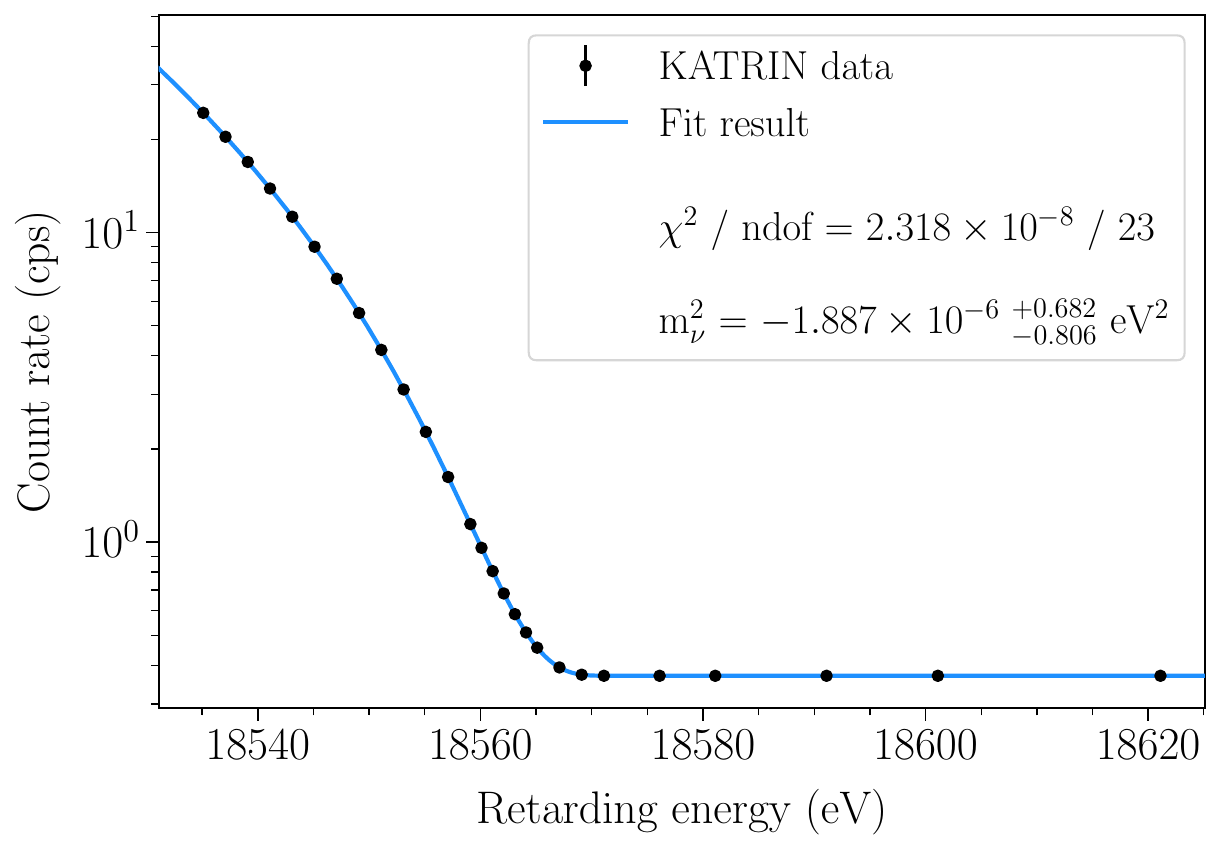}
      \caption{Illustration of the Asimov Monte Carlo data set based on the first KATRIN campaign, with corresponding four-parameter fit, yielding as expected a $\chi^2$-result close to zero as well as $m_\nu^2=\SI{0}{\square\electronvolt}$. The same pseudo-KNM1 FSD is used for the Monte Carlo data and the fit model.}
      \label{spectrum}
  \end{figure}

For the studies presented here a Monte-Carlo data set using the pseudo-KNM1 FSD introduced in detail in chap. \ref{sec:5.3} was created. This Asimov Monte-Carlo data set does not contain any statistical fluctuations and it is based on the parameters of the first KATRIN campaign. The parameters are listed in tab. \ref{fit-parameter}. A time-dependent background component of the order of $\SI{e-6}{cps \per \second}$ induced by Penning traps \cite{KNM1Paper} is neglected in this study. The spectra of all $148$ detector pixels are averaged to a single spectrum to facilitate the fits.\footnote{In the KNM1 data analysis \cite{KNM1Paper} 117 pixels were selected for the final result, yielding smaller statistics. This effect was not corrected for, since the statistical uncertainties of the fits do not impact the systematic uncertainties of these studies.}\\
During a measurement campaign, the integral $\upbeta$-decay spectrum is measured at discrete, non-equidistant retarding voltage set points which are applied repeatedly to scan the spectrum. 
The times spent at each retarding voltage set point in the Monte Carlo-generated dataset mimic the times spent during KNM1. 
Each individual scan covers the energy interval [$E_\text{0}-\SI{90}{\electronvolt}, E_\text{0}+\SI{50}{\electronvolt}]$. The set points are assumed to be perfectly reached in the simulation. A total measurement time of \SI{550}{\hour} is simulated, corresponding to \num{274} scans over the full energy range which corresponds to the total number of scans during KNM1. The generated Asimov Monte-Carlo spectrum is illustrated in fig. \ref{spectrum}.

\subsection{Fit methods\label{fit}}
The fit methods used in this work, as well as the model and fitting framework, are described in \cite{AnalysisPaper,Kleesiek:2018mel}. The Monte-Carlo data are fitted using $A_\text{s}$, $E_\text{0}$, $R_\text{bg}$ and $m_\nu^2$ as free fit parameters. In this study all other parameters in the model are fixed to the values used for generating the Asimov data set.
Like in the KNM1 analysis, the lower limit of the fit interval in this work is $E_\text{0}-\SI{40}{\electronvolt}$. The entire fit interval includes $27$ individual discrete retarding-voltage set points. With $4$ free fit parameters, this yields $23$ degrees of freedom. 
 As the Asimov Monte-Carlo data does not contain fluctuations, a fit to such data yields the exact input parameters from tab. \ref{fit-parameter} as fit results, if the same FSD is applied for generating the data and in the fit model. The fit for $m_\nu^2$ is shown in fig. \ref{spectrum}. The deviation of $m_\nu^2=\SI{-1.9e-6}{\square\electronvolt/c^4}$ from the the input value of $m_\nu^2=\SI{0}{\square\electronvolt/c^4}$ is caused by numerical inaccuracies. 
 The indicated uncertainties give an estimate of the 1-$\sigma$ statistical sensitivity of this data set.\\
 The effect of a modified FSD on $m_\nu^2$ can be investigated by fitting the Asimov data set generated using the KNM1-FSD with a model whose FSD has been modified. This mismatch between FSDs results in a shift away from the $m_\nu^2$ Monte Carlo truth of $\SI{0}{\square\electronvolt\per c^4}$, which can be used as an estimate of the impact of the FSD modification on $m_\nu^2$. The numerical deviation of \SI{2e-6}{\square\electronvolt\per c^4} mentioned above can be interpreted as the precision with which systematic uncertainties with regard to the FSDs can be determined in this study.

\section{Molecular final-states distribution (FSD)}\label{sec:FSD_theory}
\label{sec:betaspectrum}

As a result of the $\upbeta$ decay the nuclear charge of the decaying nucleus in the parent molecule (here T$_2$, HT, or DT) is increased and the formed molecular daughter ion (here $^3$HeT$^+$, $^3$HeH$^+$, or $^3$HeD$^+$, respectively) may end up in any of its molecular final states. The FSD is the distribution that describes the probabilities $\zeta_j$ with which the energy $V_j$ is left within the daughter molecular ion. 
Thus the FSD over the molecular final states enters the model of KATRIN's spectrum, eq.~(\ref{general-spectrum}), as it modifies the differential $\upbeta$-decay rate $R_\upbeta(E)$,
\begin{equation}
    R_\upbeta(E)\propto (E+m_e c^2)\cdot \sqrt{(E+m_e c^2)^2-m_e^2 c^4} \cdot \sum_j \zeta_j\cdot \epsilon_j \cdot \sqrt{\epsilon_j^2-m_\nu^2 c^4} \cdot \Theta(\epsilon_j -m_\nu c^2)\; .
\label{eq:introduction:beta_decay_fermi}
\end{equation}
Here, $m_e$ is the electron mass, $\epsilon_j=E_\text{0}-E-V_j$ is the total energy of the neutrino, and $E_\text{0}$ is the end point of the $\upbeta$-decay spectrum. Due to the (energy conserving) Heaviside step function $\Theta$ the differential decay rate $R_\beta$ is non-zero only if the kinetic energy of the neutrino ($\epsilon_j-m_\nu c^2$) is larger than or equal to $E_0-E-V_j$. At the end point the argument of $\Theta$ is equal to zero, the neutrino has zero kinetic energy, the kinetic energy $E$ of the $\upbeta$ electron has its maximum value which is equal to $E_0$, and no excitation energy is given to the molecular 
system, i.\,e.\ $V_j=0$. 
Note, in discussions of the $\upbeta$ spectrum, electron energies are generally defined relative to the end point $E_\text{0}$, thus one has {\it decreasing}  $\upbeta$-electron energies when going away from the end point. For example, a fit interval may include all $\upbeta$ electrons with energies in between $E_0$ and $E_\text{0}-E_x$. The {\it lower} limit of the fit interval is then said to lie by the energy $E_x$ {\it below} the end point. In contrast to this, the binned energies $V_j$ that are intrinsically defined to be positive\,\footnote{As is explained in more detail in the end of sec.~\ref{sec:transition_probabilities} this is only valid for a given isotopologue and initial state.} start with zero value at the end point of the $\upbeta$ spectrum and {\it increase} until the end of the fit interval. In the given example $V_j=E_x$ is said to be at the {\it upper} end of the fit interval.

In the present case of a molecular system described in Born-Oppenheimer approximation, the daughter molecular ion may be excited electronically, either to a bound state or it can be further ionised. Within every molecular electronic state, the nuclear degrees of freedom allow for rotational and vibrational excitation. In the latter case, the system may be left in a bound or an unbound (dissociative) vibrational state.
If the $^3$HeT$^+$ molecule is created in its absolute (electronic, vibrational, and rotational) ground state, the energy available to the $\upbeta$ electron and the neutrino is maximum. Until the onset of the first electronically excited state at about 19\,eV below the end point only rotational and vibrational excitation, including dissociation, are possible. In the energy range between 19\,eV and 40\,eV below the end point only electronically bound excited states of $^3$HeT$^+$ occur. Then, more than 40\,eV below the end point both, an infinite series of (bound) Rydberg states and the ionisation continuum contribute to the spectrum. The theoretical treatment of the latter is much more challenging than the one of the electronically bound states. In order to avoid the complications arising from the theoretical treatment of the ionisation continuum, the first measurement campaigns of KATRIN limited the analysis to the 40\,eV interval below the end point.

Since the initial population of the rotational states of the parent T$_2$ molecule depends on the temperature of the gas source, the FSD needs to be calculated for different initial rotational states. With the traces of HT and DT isotopologues in the gas mixture of the KATRIN source, the FSDs of HT and DT are also required for the model of the KATRIN spectrum.
To reduce the amount of data entering the KATRIN analysis and thus also the computational times when performing the fits, the FSD is typically binned and provided as a single FSD (for a given experimental campaign), which already considers the isotope mixture ($\HT, \DT, \TT$) as well as the gas temperature in the evaluation of $\zeta_j$ and $V_j$.
To increase the information content, it was proposed in \cite{PhysRevLett.84.242} to define the energy values $V_j$ as the mean transition energy for the bin $j$. The bin size should be adapted in accordance with the energy resolution of the experiment. The details on how the FSD is calculated will be discussed in the following subsections.

\subsection{Molecular transition probabilities}\label{sec:transition_probabilities}
The theoretical calculation of the FSD is based on the computation of the transition probabilities $P_{fi}^{\rm TS}$. For tritium $\upbeta$-decay this probability is given within the sudden approximation by
\begin{equation}
\centering
       P_{fi}^{\rm TS} = |\,M_{fi}^{\rm TS}\,|^2 = \left| \, \langle \Psi_{f}^{\rm HeS^{+}} | \, e^{i\boldsymbol{\tilde{K}} \cdot \boldsymbol{R}} \, | \Psi_{i}^{\rm TS} \rangle \, \right|^2 \quad .
\label{eq:molecular_fsd:molecular_matrix_element}
\end{equation}
The details of the derivation as well as the validation of the sudden approximation can be found in \cite{PhysRevC.56.2132,PhysRevC56.2162}; for a detailed description of the evaluation within the sudden approximation see \cite{valerian_paper}. In more detail, 
eq.~\ref{eq:molecular_fsd:molecular_matrix_element} describes the transition probability from a given molecular initial state $i$ of the parent molecule TS to a specific final state $f$ of the daughter molecular ion HeS$^+$ accompanying the nuclear $\upbeta$ decay. Here, the spectator S is any of the constituents in the molecule in addition to the decaying $\tritium$ atom, thus S stands in this case for $\tritium$, $\deuterium$, or $\hydrogen$ in the case of T$_2$, DT, and HT, respectively. If the final state lies within some continuum of states, {\it e.\,g.}, dissociation or ionisation, it is in fact a transition probability per unit of energy in between $E_f$ and $E_f+{\rm d}E$. The FSD $(V_j,\zeta_j)$ is obtained from the transition probabilities by a sum (or integral) of all transition probabilities $P_{fi}^{\rm TS}$ within a given energy interval. Different temperatures as they may occur in different experimental campaigns are considered by adding the transition-probability distributions of the different initial states $i$ weighted by their statistical Boltzmann distribution probability. Similarly, the transition-probability contributions of the various isotopologues are added with a statistical weight representing the relative concentrations of $\TT, \DT$, and $\HT$ for a given experimental campaign. All these summations are performed incoherently, since the occurrence of temperature and isotope distribution is a purely classical statistical effect.   

A key step is to evaluate the transition matrix element $M_{fi}^{\rm TS}$ defined in eq. \eqref{eq:molecular_fsd:molecular_matrix_element} between the wavefunction $\Psi_{f}^{\rm HeS^{+}}$ of the daughter molecular ion in state $f$ and the wavefunction $\Psi_{i}^{\rm TS}$ of the parent molecule in a given initial state $i$. All final states $f$ with excitation energies smaller than $E_\mathrm{max}$ need to be considered in the analysis of a specific neutrino-mass experiment investigating the end-point region above $E_0-E_\mathrm{max}$, {\it e.\,g.}, $E_\mathrm{max}=40$\,eV  for KATRIN KNM1. 

In eq. \eqref{eq:molecular_fsd:molecular_matrix_element} $\boldsymbol{R}$ is the vector connecting the two nuclei \footnote{The choice of the direction of the vector $\boldsymbol{R}$, {\it i.\,e.\ }whether it points towards the decaying of the spectator nucleus or not, does not change the result obtained after integration due to symmetry.} and $\boldsymbol{\tilde{K}}$ is a fractional recoil momentum, {\it i.\,e.\ }the fraction of the recoil of the emitted neutrino and the $\upbeta$-decay electron that is not transferred to the center-of-mass of the molecular ion HeS$^{+}$, but to the internal degrees of freedom. Adopting the Born-Oppenheimer approximation, the total wavefunction $\Psi$ factorises into an electronic part $\Phi$ and a nuclear-motion part $\xi$, 
\begin{equation}
    \Psi_{n,v_j,J_j,m_{J_j}}({\bf R}, {\bf r}_1, {\bf r}_2) = \Phi_n ({\bf r}_1, {\bf r}_2; R) \cdot \xi_{n;v_j,J_j,m_{J_j}} ({\bf R})  \quad. 
\label{eq:molecular_fsd:total_wavefunction}
\end{equation}

Here $R=|\bf{R}|$ is the internuclear separation (distance between the two nuclei), ${\bf r}_j$ ($j=1,2$) are the coordinates of the two electrons , $J_j\,$ the total angular momentum, $m_{J_j}$ its projection along the z-axis, $v_j$ the vibrational quantum number and $n$ the electronic state of the molecule. The electronic wavefunctions $\Phi$ have only a parametric dependency on $R$, i.\,e.\ $R$ is kept constant when solving the eigenvalue equation 
\begin{equation}\label{eq:electronic_eigenvalueproblem}
    \hat{\rm H}_{\rm el} \, \Phi_n ({\bf r}_1, {\bf r}_2; R) = V_n^{\rm BO}(R) \: \Phi_n ({\bf r}_1, {\bf r}_2; R)
\end{equation}
with the electronic Hamiltonian
\begin{equation}
    \hat{\rm H}_{\rm el} = -\,\frac{\hbar^2}{2 m^2_e} \: \left(\nabla_1^2 + \nabla_2^2\right) - \, \frac{e^2}{4\,\pi\epsilon_0} \left( \frac{Z_\text{A}}{r_{1\text{A}}} - \frac{Z_\text{A}}{r_{2\text{A}}} - \frac{Z_\text{B}}{r_{1\text{B}}} - \frac{Z_\text{B}}{r_{2\text{B}}} + \frac{1}{r_{12}} + \frac{Z_\text{A} Z_\text{B}}{R}\right)\;.
\label{eq:molecular_fsd:electronic_hamiltonian}
\end{equation}
Here the $V_n^{\rm BO}$ are the internuclear-separation dependent electronic energies for electronic state $n$, the so-called Born-Oppenheimer potential curves. A and B denote the nuclei, 1 and 2 the electrons, $r_{ij} = |{\bf r}_i - {\bf r}_j|$, and $Z_A$ ($Z_B$) is the charge number of nucleus A (B). 

The substitution of eq.\,\eqref{eq:molecular_fsd:total_wavefunction} into eq. \eqref{eq:molecular_fsd:molecular_matrix_element}, considering only the decay of molecules that were initially in the electronic ground state ($n_i=0$), and adopting the notation $n_f\equiv n$ yields the volume integral with respect to the nuclear-separation coordinate $\boldsymbol{R}$ 
\begin{equation}
    M_{n,f;i}^{\rm TS}(\tilde{K}) \:= \: \iiint\limits_{-\infty}^{+\infty} \left\{\xi_{n;v_f,J_f,m_f}^{\rm HeS^{+}}(\boldsymbol{R}) \right\}^*  \; S_n(R) \cdot e^{i \boldsymbol{\tilde K} \; \boldsymbol{R}} \; \xi_{0;v_i,J_i,m_i}^{\rm TS}(\boldsymbol{R}) \;  {\rm d}V_R 
\label{eq:molecular_fsd:molecular_matrix_element_separation}
\end{equation}
with the electronic transition overlap matrix element $S_n(R)$ defined as the six-dimensional integral 
\begin{equation}
    \begin{split}
        S_n (R) &=  \iiint\limits_{-\infty}^{+\infty}\iiint\limits_{-\infty}^{+\infty} \left\{\Phi_n^{\rm HeS^+}({\bf r}_1, {\bf r}_2; R)\right\}^*\,\Phi_0^{\rm TS}({\bf r}_1, {\bf r}_2; R)\; \text{d}{V}_1 \text{d}{V}_2 \quad .
    \label{eq:molecular_fsd:electronic_overlap}
    \end{split}
\end{equation}
The nuclear-motion wavefunctions $\xi$ describing the rotational $J_j\, , m_{J_j}$ and vibrational $v_j$ degrees of freedom of the molecule, i.\,e.\ all but the translational degree of nuclear motion, are the solutions of the eigenvalue equation 
\begin{equation}
    \hat{\rm H}_{\rm nuc}^{(n)} \; \xi_{n;v_j,J_j,m_{J_j}} ({\bf R}) 
        = E_{n;v_j,J_j,m_{J_j}} \;\xi_{n;v_j,J_j,m_{J_j}} ({\bf R}) 
\end{equation}
with the nuclear-motion Hamiltonian 
\begin{equation}
    \hat{\rm H}_{\rm nuc}^{(n)} = -\frac{\hbar^2}{2m_{\mu}} \,\nabla_{R}^2 + V_{n}^{\rm BO}(R) \quad .
\label{eq:molecular_fsd:nuclear_hamiltonian}
\end{equation}
Here $m_{\mu}$ is the reduced mass of the two atoms T and S, which will be discussed in more detail at the end of sec.~\ref{sec:deliberate_approximations}) and in sec.~\ref{reduced-mass}. Since $V_n^{\rm BO}$ is only a function of the absolute value of the internuclear distance, the angular solutions of eq. \eqref{eq:molecular_fsd:nuclear_hamiltonian} are simply the spherical harmonics $Y$, 
\begin{equation}
    \xi_{n;v_j,J_j,m_{J_j}} ({\bf R}) = \frac{\chi_{n;v_j,J_j} (R)}{R} \; Y_{J_j,m_{J_j}} (\theta_R,\varphi_R) \quad , 
\label{eq:chi}      
\end{equation}
while the radial wavefunctions $\tilde{\chi}=\chi \, R^{-1}$ are solutions of the eigenvalue equation
\begin{equation}
    \hat{\rm H}_{{\rm nuc},J}^{(n)} \, \tilde{\chi}_{n;v_j,J_j} (R) 
        = E_{n;v_j,J_j} \, \tilde{\chi}_{n;v_j,J_j} (R) 
\label{eq:nuclear_eigenvalue_problem}
\end{equation}
with the radial nuclear-motion Hamiltonian  
\begin{equation}
    \hat{\rm H}_{{\rm nuc},J}^{(n)} \;=\; -\: \frac{\hbar^2}{2m_{\mu}} \: \frac{\mathrm{d}^2}{\mathrm{d}R^2}  + \frac{\hbar^2}{2m_{\mu}} \: \frac{J(J+1)}{R^2} \:+\: V_{n}^{\rm BO}(R) \quad .
\label{eq:molecular_fsd:radial_hamiltonian}
\end{equation}
As a consequence of the spherical symmetry, the energy $E_{n;v_j,J_j}$ does not depend on the magnetic quantum number $m_{J_j}$.  
  
The separation of the wavefunctions $\xi$ into an angular and a radial part together with an expansion of  the $\exp(i\boldsymbol{\tilde{K}}\boldsymbol{R})$ term in a series of spherical harmonics times spherical Bessel functions $j_\ell(\tilde{K} R)$ allows for a straightforward integration over the angles by using the properties of the spherical harmonics (leading to terms that can be expressed using Wigner 3j symbols or Clebsch-Gordan coefficients), for details see \cite{valerian_paper}. The remaining task is the calculation of one-dimensional integrals over the radial coordinate $R$, 
\begin{equation}
    I_{0;v_i.J_i}^{n;v_f,J_f} (\tilde{K}) \; = \; \int\limits_{0}^{\infty} \left\{\tilde{\chi}_{n;v_f,J_f}^{\rm HeS^{+}}(R) \right\}^*  \; S_n(R)\; j_\ell (\tilde{K} R) \; \tilde{\chi}_{0;v_i,J_i}^{\rm TS}(R) \:  {\rm d}R
      \quad .
\label{eq:molecular_fsd:radial_integrals}
\end{equation}
  
The evaluation of the transition matrix elements $M_{fi}$ thus splits basically into four steps. First, the solution of the electronic eigenvalue problem eq. \eqref{eq:electronic_eigenvalueproblem} is required in order to obtain the electronic wavefunctions $\Phi_n$ and the Born-Oppenheimer potential curves $V_n^{\rm BO}(R)$. The former are needed for the calculation of the electronic transition overlaps $S_n$ according to eq. \eqref{eq:molecular_fsd:electronic_overlap} and the latter as an input for the radial nuclear-motion eigenvalue problem eq. \eqref{eq:molecular_fsd:radial_hamiltonian}. Second, the electronic overlaps $S_n(R)$ are calculated according to eq. \eqref{eq:molecular_fsd:electronic_overlap}. Third, the wavefunctions of nuclear motion $\chi$ and the complete molecular (rotational, vibrational, and electronic) energies $E_{n;v_j,J_j}$ are obtained by solving eq. \eqref{eq:nuclear_eigenvalue_problem}. Fourth, the matrix elements $M_{fi}$ in eq.\eqref{eq:molecular_fsd:molecular_matrix_element_separation} are evaluated adopting the corresponding Clebsch-Gordan algebra and calculating the integrals $I$ in eq. \eqref{eq:molecular_fsd:radial_integrals} by means of numerical quadrature.  
  
At the temperature of 30\,K in the first KATRIN science campaign mostly rotational and some vibrational excitation of the parent molecule is possible, while the electronic excitation is negligible and thus $n_i=0$ is used. In the case of a homonuclear diatomic parent molecule like $\TT$ also the spin statistics (ortho, para) needs to be taken into account, since the symmetry of the nuclear spins allows only for either even or odd values of the rotational quantum number $J$. When adding the binned FSDs for different initial states and isotopologues of the parent molecules one needs to adjust the energy scales (energy offsets).\,\footnote{Since any excitation energy left in the molecular system is not available to the $\upbeta$ electron, the maximum energy of the $\upbeta$ electron, which is the end point $E_0$ of the $\upbeta$ spectrum, occurs, if the molecular daughter ion is generated in its ground state. More accurately, the end-point energy $E_0$ depends on the energy change within the molecular system and thus the difference between the energies of the initial state of the parent molecule and the absolute (electronic, vibrational, and rotational) ground state of the daughter molecular ion. This energy difference depends on the isotopologue, but also on the initial state. The FSDs obtained individually for specific initial (rotational) states and isotopologues are all adjusted to a common energy scale set by the $E_0$ value that corresponds to a $\upbeta$ decay of the T$_2$ isotopologue in its ground state, the most common case in the KATRIN experiment.} 
  
The calculation is performed for all isotopologues, {\it i.\,e.\ }steps 3 and 4 as well as the binning (including the temperature effects) are repeated for $\TT$, $\DT$, and $\HT$. The three binned spectra are combined by first adjusting the different energy scales and then adding the probabilities with the corresponding weights determined by relative isotopologue concentration. Consequently, also new mean transition energies per bin are obtained. This whole procedure results finally in the FSD (a list of ($V_j$,$\zeta_j$) values) used in the fit of $m_\nu^2$. 
Once the transition matrix elements $M_{f,i}^{\rm TS}(\tilde{K})$ (see eq.~\eqref{eq:molecular_fsd:molecular_matrix_element}) for the three isotoplogues and a sufficient number of initial rotational states are evaluated and stored, the FSDs for different bin sizes, temperatures, or isotopologue mixtures can be obtained straightforwardly. 
The molecular calculations are performed adopting atomic units in version Hartree (setting $\hbar=m_e=4\pi\epsilon_0=e=1$) with distances given in units of Bohr, $a_0\approx 5.29\cdot 10^{-11}\,$m, energies in Hartree, $E_H = \hbar^2/(m_e a_0^2) \approx 27.2114\,$eV, and masses in units of the mass of the electron, $m_e\approx5.11\cdot 10^5$\,eV/c$^2$.  
  
These units are used in the following if not stated otherwise.

%
% POTENTIAL CURVES AND ELECTRONIC OVERLAPS
%
\subsection{Electronic potential curves and overlaps\label{sec:potentialcurves}}    
There are various ways to find the solutions of the electronic eigenvalue problem eq.\eqref{eq:electronic_eigenvalueproblem} for diatomic two-electron molecules. In order to obtain efficiently highly accurate solutions, the wavefunctions $\Phi$ may be expressed as a linear combination of $N$ explicitly correlated two-electron basis functions $\phi$, so called geminals, 
\begin{equation}
    \Phi_n(\boldsymbol{r}_1,\boldsymbol{r}_2; R) \;=\; \sum_{j=1}^N \: c_{n,j} (R) \;   
    \phi_j(\boldsymbol{r}_1,\boldsymbol{r}_2; R) \quad .
\label{eq:Phi_n}                                                
\end{equation}
These geminal expansions converge much faster than the most accurate standard quantum-chemist\-ry approach known as full configuration interaction (CI) (also known as exact diagonalisation, see \cite{szaboostlund}). In the CI approach, the basis functions $\phi$ are uncorrelated tensor products (more accurately properly anti-symmetrised Slater determinants) of two one-electron basis functions, usually Hartree-Fock orbitals. 
    
The (non-relativistic) one-electron diatomic problem is separable in prolate spheroidal coordinates. For an electron $i$ these coordinates are defined as $\xi_i=(r_{i,A}+r_{i,B})/R$, $\eta_i=(r_{i,A}-r_{i,B})/R$, and the angle $\phi_i$ around the internuclear axis, where A and B are the two foci of the ellipse, naturally to be chosen as the positions of the two nuclei. The so-called Ko{\l}os-Wolniewicz (or James-Coolidge) basis functions (see, e.\,g., \cite{dia:kolos85}) are defined as  
\begin{equation}
    \phi_{j,\text{KW}} (\boldsymbol{r}_1,\boldsymbol{r}_2; R) \;=\; {\cal N}_j \: \rho_{12}^{\mu_{j}}  \: \xi_1^{\lambda_{j}} \: \xi_2^{\bar{\lambda}_{j}} \: \eta_1^{\nu_{j}} \: \eta_2^{\bar{\nu}_{j}} \; e^{- \alpha \xi_1 - \bar{\alpha} \xi_2 - \beta \eta_1 - \bar{\beta} \eta_2 } .
    \label{eq:molecular_fsd:kolos_basis_functions}
\end{equation}
Here, $\rho_{12}=|\boldsymbol{r}_1-\boldsymbol{r}_2|$ and ${\cal N}_j$ is a prefactor that depends on $R$, but is constant for a specific value of $R$.\footnote{In literature, different choices for $\rho_{12}$ and thus also $N_j$ are in use.} 
Furthermore, $\alpha$, $\bar{\alpha}$, $\beta$, and $\bar{\beta}$ are (in general) non-integer parameters that form a {\it base}. This base is identical for all basis functions of a given basis set. A specific basis function $j$ for a given base is characterised uniquely by a set of five integers, the quintuple $\{\mu_j, \lambda_{j}, \bar{\lambda}_j, \nu_j, \bar{\nu}_j\}$. More accurately, linear combinations of the basis functions in eq. \eqref{eq:molecular_fsd:kolos_basis_functions} are adopted depending on the symmetry. A {\it basis set} is thus defined by the specification of a base and a set of quintuples, one quintuple per basis function. A summary of the namings related to the base used in this paper is given in app. \ref{glossary}.
    
Since the basis functions defined in eq. \eqref{eq:molecular_fsd:kolos_basis_functions} are not orthogonal to each other, their use in the {\it ansatz} eq. \eqref{eq:Phi_n} for solving the electronic eigenvalue eq. \eqref{eq:electronic_eigenvalueproblem} yields the generalised eigenvalue problem  
\begin{equation}
    {\bf H}_{\rm el} \: {\bf c}_n(R) \;=\; V_n^{\rm BO}(R) \: {\bf\tilde{S}} \: {\bf c}_n
\label{eq:molecular_fsd:generalized_eigenvalue_problem}
\end{equation}
with the electronic Hamiltonian matrix ${\bf H}_{\rm el}$ with the matrix elements
\begin{equation}
    h_{i,j}(R) \;=\; \langle\,\phi_i(R)\,|\,\hat{\rm H}_{\rm el}\,|\,\phi_j(R) \,\rangle
\label{eq:hij}    
\end{equation}
and the overlap matrix ${\bf\tilde{S}}$ with the matrix elements
\begin{equation}
    \tilde{s}_{i,j}(R) \;=\; \langle\,\phi_i(R)\,|\,\phi_j(R) \,\rangle \quad .
\label{eq:tilde_sij}    
\end{equation}
In eq. \eqref{eq:hij} and eq. \eqref{eq:tilde_sij} the braket notation implies an integration over the coordinates of the two electrons, only. The solution of the generalised eigenvalue problem eq. \eqref{eq:molecular_fsd:generalized_eigenvalue_problem} for different values of the internuclear separation $R$ provides the Born-Oppenheimer potential curves $V_n^{\rm BO}(R)$ that enter the corresponding nuclear-motion Hamiltonian $\hat{\rm H}_{{\rm nuc},J}^{(n)}$ eq. \eqref{eq:molecular_fsd:radial_hamiltonian} and the eigenvector coefficients ${\bf c}_n(R)$ that define the electronic wavefunctions according to eq. \eqref{eq:Phi_n}. Solving this equation for both, the parent molecule TS and the daughter molecular ion HeS$^+$, allows finally for the evaluation of the electronic transition matrix elements (asymmetric overlaps) $S_n(R)$ according to eq. \eqref{eq:molecular_fsd:electronic_overlap} as
\begin{equation}
    S_n(R) \;=\; \sum\limits_{i=1}^{N^{\rm HeS^+}} \:  \sum\limits_{j=1}^{N^{\rm TS}} \;
                c_{n,i}^{{\rm HeS}^+} (R) \: 
                s_{i,j}(R) \:
                c_{0,j}^{\rm TS}(R)
\end{equation}
with the matrix elements
\begin{equation}
    s_{i,j}(R) \;=\; \langle\,\phi_i^{{\rm HeS}^+}(R)\,|\,\phi_j^{\rm TS}(R) \,\rangle \quad .
\label{eq:sij}    
\end{equation}
The indices denoting the molecular system emphasise that usually the number of basis functions $N$ and the base are chosen differently for parent and daughter molecules as to optimally describe these different systems in a more efficient way. The potential curves and electronic overlaps used in the KNM1-FSD calculation were obtained with a variant of a code originally written by Ko{\l}os, Wolniewicz and co-workers (for brevity abbreviated as Ko{\l}os code in the following).\footnote{The full history of the code cannot be recovered. Its origin certainly dates back at least to the beginning of the 1980s and the basis-set definition used corresponds to the one given in \cite{dia:kolo66}, but it was modified by various authors since then, including the implementation of a (slightly) more automatic memory management by A.~Saenz.}
    
\subsection{Basis-set convergence parameter $\Omega$ and number of included final states}\label{omega-variation}
According to the variational principle, the energies (and eigenvectors) obtained in a basis-set calculation can be systematically improved by increasing the basis set, {\it i.\,e.\ }by the addition of more and more basis functions. However, an increase of the size of a basis set typically causes the contained set of basis functions to become linearly dependent due to the finite numerical precision. 
Instead of canonical orthogonalisation that is of limited use, another solution is provided by using variable precision in the calculation, {\it i.\,e.\ }the number of digits used in the representation of the floating point numbers. Based on a library (MPFUN) that allows for the use of flexible precision with FORTRAN programs, Pachucki, Zientkiewicz, and Yerokhin \cite{CPC} implemented a new code (H2SOLV), which uses Ko{\l}os-Wolniewicz basis functions but allows for the systematic increase of the basis set. More accurately, Pachucki {\it et al.\ }use the slightly different notation and form
\begin{equation}
\label{eq:molecular_fsd:pachucki_basis_functions}
    \phi_{j,{\rm Pa}} = e^{- u \xi_1 R - w \xi_2 R - y \eta_1 R - x \eta_2 R} \rho_{12}^{n_{0,j}} \eta_1^{n_{1,j}} \eta_2^{n_{2,j}} \xi_1^{n_{3,j}} \xi_2^{n_{4,j}} \quad , 
\end{equation}
compared to the basis function in eq.
\eqref{eq:molecular_fsd:kolos_basis_functions}. The difference is the explicit inclusion of the internuclear separation $R$ in the non-integer base parameters $u,w,y$, and $x$. For all calculations in this work a modified version of the H2SOLV code, in the following called H2SOLVm, was used.

With the linear-dependency problem under control (though paying the price of requiring much larger computational resources), the basis set can be systematically increased by simply adding more and more basis functions $j$ that differ only by the quintuples, which are denoted by $\{n_{0,j}, n_{1,j}, \\ n_{2,j}, n_{3,j}, n_{4,j}\}$ in the H2SOLV code. In order to systematically increase the basis set, the H2SOLV code introduces a single parameter $\Omega$ that, together with the base, {\it i.\,e.\ }the values of $u,w,y$, and $x$, defines the complete basis set. For a given value of $\Omega$ all basis functions $j$, i.\,e.\ all quintuples $\{n_{0,j}, n_{1,j}, n_{2,j}, n_{3,j}, n_{4,j}\}$ with integer values $n_{i,j}$, are included that fulfill  
\begin{equation}
    \sum_{i = 0}^{4} n_{i,j} \;\leq\; \Omega .
\label{eq:molecular_fsd:omega}
\end{equation}
In more detail, symmetries (especially spin symmetry as well as D$_{\infty h}$ and C$_{\infty v}$ symmetries\footnote{$D_{\infty h}$ is the point group of the electronic Hamiltonian of a homonuclear diatomic molecule like H$_\text{2}$ with inversion symmetry. $C_{\infty v}$ is the point group in the absence of inversion symmetry, here for the electronic Hamiltonian for all heteronuclear diatomic molecules like HeH$^+$ (or other linear molecules without inversion symmetry).} for a homonuclear or heteronuclear diatomic molecule, respectively) are explicitly considered. Therefore, corresponding linear combinations of the basis functions eq. \eqref{eq:molecular_fsd:pachucki_basis_functions} are used that transform like irreducible representations of the corresponding symmetry group. As a consequence, the total number of basis functions $N(\Omega)$ is smaller than the value obtained from the restriction \jannis{eq.}\eqref{eq:molecular_fsd:omega}. Due to the variational principle, the wavefunction is improved (or remains of equal quality\footnote{\AS{A higher accuracy of energies and wavefunctions is, due to the variational principle, equivalent to a higher quality.}}), if more basis functions are added. Therefore, $\Omega$ is the central convergence parameter. An exact solution \jannis{will} be obtained, if $\Omega$ approaches infinity. \jannis{As shown in sec. \ref{sec:potential_curves}, if $\Omega$ is increased in a set of test FSDs, a well-defined convergence behaviour of the extracted $m_{\nu}^2$ is observed.}
    
It should be noted that the basis-set optimisation based only on the $\Omega$ variation is less efficient than the careful selection of basis functions (quintuples), as it was done for the previous FSD calculations \cite{Fackler1985,PhysRevC.73.025502,PhysRevLett.84.242}, including the KNM1 FSD. However, especially for the present purpose of an uncertainty investigation it is advantageous to have a single (or only a few) convergence parameter(s). 
    
As already mentioned, all final states need to be considered in the FSD that have energies lying in the energy interval used in the KATRIN analysis. 
\jannis{Even within the 40\,eV \ChWe{wide} energy interval \ChWe{below the end point} considered in this work, we must consider the infinite number of electronically bound states: the Rydberg states.} On the other hand, if a finite basis set comprising $N$ basis functions is used, only a finite number of $N$ states can be obtained. Furthermore, depending on the chosen base, the fraction of these states representing bound or discretised continuum states differs, even for the same value of $\Omega$. More importantly, since the convergence study is based on a systematic  enlargement of $\Omega$, the size of the basis-set and thus the number of basis functions $N$ increases. As a consequence, the number of electronic states obtained within the 40\,eV energy interval increases with $\Omega$, and this in a rather unpredictable way. While this would be already an issue for atoms, in the case of molecules the nuclear motion leads to the potential curves (see eq. \eqref{eq:electronic_eigenvalueproblem} and eq. \eqref{eq:molecular_fsd:nuclear_hamiltonian}). In principle, any electronic state $n$ may contribute to the 40\,eV interval, if its energy $V_n^{\rm BO}(R)$ lies for some value of the internuclear separation $R$ below 40\,eV. In summary, the number of electronic states that is included in the generation of the test FSDs needs to be adapted, if $\Omega$ is varied, in order to obtain consistent FSDs covering the same energy interval. Consequently, and keeping in mind that all the electronically excited states of $\HeH$ are purely repulsive in the Franck-Condon window see \cite{valerian_paper,brandsenjoachin}, the Born-Oppenheimer potential curves for all states obtained with a given basis set are computed. \AS{Then the number of states, $N_\text{states}$, are determined that lie below the $\SI{40}{\electronvolt}$ threshold, i.\,e.\ the energy difference to the absolute (electronic, vibrational, and rotational) ground state is less than 40\,eV. In this selection of states the maximum distance considered for the calculation of the excited states is set to $R_\text{max, ex} = 4\,\text{a}_0$.} The resulting value of $N_\text{states}$ for each $\Omega$ and thus included in the corresponding FSD calculation is listed in tab. \ref{tbl:molecular_fsd:n_states}. (Note, in the calculation of the KNM1 FSD 13 electronically bound states were considered, as obtained with the basis set used in that calculation.) Since the adiabatic corrections (see \ref{corrections}) are not considered for the electronically excited states, this leads to a slight overestimation of $N_\text{states}$. These additional states will result in additional probability in the FSD, but it will only affect the FSD \jannisfourty{more than} 40\,eV \jannisfourty{below the end point,} \jannis{so they do not need consideration here.}

\begin{table}[htpb]
    \centering
    \caption{
        Number of electronic states $N_\text{states}$ below $\SI{40}{\electronvolt}$ used for the computation of the FSD in dependence of $\Omega$.
    }
    \begin{tabular}{c|c|c|c|c|c|c|c}
        \toprule
        $\Omega$          & 4 & 5  & 6  & 7  & 8  & 9  & 10 \\
        \midrule
        $N_\text{states}$ & 6 & 10 & 13 & 18 & 24 & 31 & 36 \\
        \bottomrule
    \end{tabular}
    \label{tbl:molecular_fsd:n_states}
\end{table}

\subsection{Nuclear motion and transition matrix elements\label{sec:nuclear-motion}}

For every electronic state ($n_i=0$ for the electronic ground state of TS; $n_f$ for the electronic ground and excited states of $^3$HeS$^+$, S=H, D, or T) the radial nuclear Schrödinger eq. \eqref{eq:nuclear_eigenvalue_problem} has to be solved separately using the corresponding Born-Oppenheimer potential curve $V_n^{\rm BO}$ of that electronic state. This yields the radial part $\tilde{\chi}_{n;v,J}$ of the nuclear-motion wavefunction and the complete molecular energy $E_{n;v,J}$ of the electronic, vibrational, and rotational state within the Born-Oppenheimer approximation. While the electronic ground states of parent and daughter molecules support a number of bound solutions besides states within the dissociation continuum, the potential curves of the electronically excited states are almost purely repulsive, {\it i.\,e.\ }they do not support bound molecular states, but only dissociative ones; at least within the considered $R$ range. Clearly, due to the dissociation continua there is an infinite number of nuclear-motion states for each potential curve, and thus a selection is needed in practice. 
    
There are two basic types of approaches \jannis{for solving} the eigenvalue equation \eqref{eq:nuclear_eigenvalue_problem} and \jannis{thus for finding} the nuclear-motion wavefunctions and total molecular energies within the Born-Oppenheimer approximation\jannis{: direct numerical integration or the variational (basis-set) approach.} Since it is a one-dimensional differential equation, a numerical solution by integration is in principle straightforward. Bound states are obtained via the shooting method where one bound state is found at a time. For accurate and numerically stable solutions often the method described by Cooley \cite{Cooley61}, a Numerov integration, is used. This approach was adopted in the FSD calculations in \cite{Fackler1985} and in \cite{PhysRevLett.84.242}. The electronic ground state of $^3$HeS$^{+}$ supports not only a large number of rovibrational \jannis{(rotational and vibrational)} bound states and a smooth dissociative continuum, but in this dissociative continuum there occur a large number of predissociative resonances, {\it i.\,e.\ }metastable states leading to pronounced structures in transition spectra. The very narrow resonances behave almost like (non-dissociative) bound states embedded in the background continuum of states. Those states can be obtained by treating them as bound states with the mentioned shooting method combined with Numerov integration. The broader metastable and, of course, the purely dissociative states are scattering states and thus their proper treatment requires an energy (or momentum) $\delta$-function normalization instead of the bound-state Kronecker-$\delta$ normalization. 
    
Alternatively to the one-dimensional numerical integration, the radial nuclear-motion eigenvalue problem may be solved by adopting a basis-set expansion similar to the one that was described in sec. \ref{sec:potentialcurves} for the solution of the two-electron problem. \jannis{M}uch simpler basis functions are needed in this one-dimensional case \jannis{, so} the convergence is much faster. One popular and flexible choice of basis functions are $B$ splines, see, e.\,g.\ \cite{deBoor}, that are a generalization of cubic splines to arbitrary polynomial order $k$, 
\begin{equation}
    \tilde{\chi}_{n;v,J} \;=\; \sum\limits_{n_B=1}^{N_B} \: d_{n;v,J;n_B} \: B_{n_B}^{(k)}(R) \quad .
\end{equation}
This {\it ansatz} inserted into the eigenvalue equation leads to a generalised, but sparse eigenvalue problem, since the $B$ splines are non-orthogonal and only overlap with a small number of neighbor $B$ splines. The bandwidth of the sparse matrix depends directly on the adopted order $k$ of the $B$ splines. Besides their order, the $B$-spline basis is defined by the so-called knot sequence, {\it i.\,e.\ }the sequence of radial grid points defining their support. The radial grid is finite and thus the wavefunctions are confined within some finite (spherical) box with radius $R_{\rm max}$. Fixed boundary conditions are applied at the upper grid boundary (setting the wavefunction and its derivatives to zero at the last grid point) yielding discretised states in the  dissociative continuum with Kronecker $\delta$ normalization. For convenience, the wavefunction and its derivative at $R=0$ are set to zero. These boundary conditions are easily implemented by removing the corresponding $B$ splines. Therefore, if $N_B$ is the number of adopted $B$ splines  the number of states that can be obtained is $N_B-2$. While the density of the knot points is decisive for the state with maximum energy that is yielded, the box size and thus $R_{\rm max}$ determines the density of states in the resulting discretised dissociation continuum. Since it is a single-channel scattering problem, a renormalisation to energy-normalised states is, in contrast to the case of the ionization continuum of two electrons, straightforward and thus transition-probability densities (per unit of energy) can easily be obtained from the discretised transition probabilities (using a renormalisation procedure based on the energy density of states). 
    
Once the radial nuclear-motion wavefunctions $\tilde{\chi}$ are obtained in either way, they are then used (together with the electronic transition matrix elements $S_n$ and the spherical Bessel functions) to solve the corresponding radial integrals $I$ , see eq. \eqref{eq:molecular_fsd:radial_integrals}. This integration is done numerically using quadrature\AS{. This quadrature is exact in the case of $B$ splines}, if Gaussian quadrature is adopted and the product of the spherical Bessel function and the electronic transition matrix elements can be expressed as a finite polynomial. The integrals finally yield the state-to-state transition probabilities $P$ according to eq. \eqref{eq:molecular_fsd:molecular_matrix_element}.

\section{Quantifying the FSD uncertainty}\label{sec:quantifying}
The calculation steps for the FSD as described in sec. \ref{sec:FSD_theory} introduce uncertainties on the resulting shape of the FSD and therefore on the neutrino mass extracted from a measured $\upbeta$-decay spectrum. In the following, the different kinds of uncertainties which contribute to the FSD uncertainty are introduced. Afterwards, it is described how the total FSD uncertainty had previously been estimated for the first two measurement campaigns of KATRIN.

\subsection{Types of uncertainty contributions from the FSD\label{types}}
There are different types of uncertainties of the FSD obtained from theoretical calculations, four with origin in the theory and calculation itself, and one with experimental origin. 
First of all, there are uncertainties due to the adopted approximations (sec. \ref{sec:deliberate_approximations}), {\it e.\,g.\ }the sudden approximation. Assessing the validity of the approximations ensures that the uncertainty induced by them is sufficiently small. Second, there are possible uncertainties in constants entering the calculation (sec. \ref{sec:conversion_factors}), for example the nuclear masses. Third, there are uncertainties due the use of finite basis-sets and finite numerical precision in the {\it ab initio} calculations of energies, wavefunctions, transition matrix elements, and probabilities (sec. \ref{sec:convergence_uncertainty}). 
Since some of the corrections to the adopted approximations are calculated using the calculated energies and wavefunctions the evaluated corrections themselves have uncertainties due to the adopted finite basis sets. Finally, there is also a possibility of errors made in the computer codes or the input files (sec. \ref{sec:code_input_errors}).  
    
If the FSD is binned and isotopologue composition as well as temperature (including possible spin-statistical effects) are considered, {\it i.\,e.\ }if the FSD is a sum of FSDs with corresponding statistical weighting factors, an additional uncertainty is introduced. However, in this case the uncertainty has a purely experimental origin, since it is caused by the uncertainty in temperature and isotopologue distribution in a given experiment. Estimations on the effect of these uncertainties with experimental origin on $m_\nu^2$ can be found in sec. \ref{subsec:binning} and app. \ref{app:temperature}.

\subsubsection{Approximations adopted in the FSD calculation}\label{sec:deliberate_approximations}
The FSDs used so far in the analysis of tritium $\upbeta$-decay experiments rely on the sudden approximation.
Near the end point of the $\upbeta$-decay spectrum, the  $\upbeta$-decay electron is emitted with large kinetic energies of more than 18\,keV and thus with very high speed. The remaining atomic or molecular system experiences in first order the decay process as a sudden change of the charge of the decaying nucleus. 
The validity of the sudden approximation including an explicit calculation of the leading corrections was investigated in \cite{PhysRevLett.77.4724,PhysRevC.56.2132,PhysRevC56.2162}. It was found that the leading correction stems from the Coulomb distortion of the wavefunction of the emitted $\upbeta$-decay electron by the decaying nucleus and this effect is described by a simple (and well-known) factor, the so-called Fermi function. The function depends on the kinetic energy of the $\upbeta$-decay electron and the nuclear charge of the decaying nucleus. This factor is included in the spectrum model of KATRIN (as was also done in the analysis of previous experiments). 
The next largest correction to the sudden approximation due to the spectator nucleus and the two molecular electrons is explicitly given in \cite{PhysRevC56.2162} where it was found that it is mostly proportional to the probability in the sudden approximation. Since the $m_\nu^2$ measurement by KATRIN does not depend on the total probability, but the relative distribution of the probabilities as a function of energy, the effective impact of the correction to the sudden approximation is practically reduced by about one order of magnitude \cite{PhysRevC56.2162}. Based on these results it was concluded that at least for the first science campaigns of KATRIN the sudden approximation is applicable, see sec.~\ref{subsec:suddenapprox} for more details. 
  
In the calculation of the KNM1 FSD (as in the ones before) the Born-Oppenheimer approximation is applied. The electronic problem in eq.~\ref{eq:electronic_eigenvalueproblem} is solved as a function of a fixed internuclear separation yielding potential curves as described in sec.\,\ref{sec:potentialcurves}. The electronic wavefunctions are then used in the calculation of the electronic transition matrix elements $S_n$. The potential curves enter the Schr\"odinger equation describing the vibrational and rotational degrees of freedom of the nuclei. 
In the derivation of the Born-Oppenheimer approximation, the matrix elements $\langle\,\Phi_n\,|\,\Delta_R\,|\,\Phi_m\,\rangle$ stemming from the action of the kinetic-energy operator of the nuclei $\Delta_R$ on the electronic wavefunctions $\Phi$ are neglected. Considering the diagonal matrix elements (the ones between the same electronic wavefunction, $n=m$) leads to the adiabatic approximation, while the off-diagonal matrix elements ($n\neq m$) that couple different electronic Born-Oppenheimer states are known as non-adiabatic corrections. For the ground states of both the parent and the daughter molecules the adiabatic corrections are included in the KNM1 FSD (as they were in the FSDs in \cite{Fackler1985} and \cite{PhysRevLett.84.242}). The non-adiabatic corrections have been considered in the calculation of the transition probabilities to a variety of rovibrational states of $\HeT$ in \cite{dia:nonadiabatic}. However, it was observed that although the transition probabilities to some individual states changed significantly, transition probabilities are effectively only shifted within a very small energy window, leading to variations in the binned FSD of the order of $\mathcal{O}(10^{-4})$. The effect of this correction is similar to the effect of the binning in the case of the KNM1 FSD, i.\,e.\ of the order of 10 to 100\,meV. Thus the non-adiabatic corrections can be neglected for the first science campaigns of KATRIN.            
  
The calculation of the FSD was performed using basically non-relativistic quantum mechanics and neglecting effects stemming from quantum electrodynamics. Only for the electronic ground state of the parent molecule first-order relativistic and radiative corrections were included, since they were available in literature \cite{dia:woln93}. All of the corrections (adiabatic $V_n^{\rm ad}$, relativistic $V_n^{\rm rel}$ and radiative $V_n^{\rm rad}$) are included by adding them to the corresponding Born-Oppenheimer potential curve $V_n^{\rm BO}$, resulting in the corrected potential curve
\begin{equation}
    V_n^{\rm cor} = V_n^{\rm BO} + V_n^{\rm ad} + V_n^{\rm rel} + V_n^{\rm qed}\quad .
    \label{eq:corrected_potential}
\end{equation}
The corrected potential curve $V_n^{\rm cor}$ is used, instead of the Born-Oppenheimer potential curve $V_n^{\rm BO}$, in the nuclear-motion Hamiltonian from eq. \eqref{eq:molecular_fsd:radial_hamiltonian}. Since the Born-Oppenheimer potential curves and corrections are not given for the same internuclear separations $R$, the corrections were spline interpolated and then evaluated at the internuclear separations of the Born-Oppenheimer potential curves. The effect of the corrections to the Born-Oppenheimer approximation on $m_\nu^2$ is given in sec. \ref{corrections}.

Another approximation has been adopted in order to obtain the operator $e^{i\boldsymbol{\tilde{K}}\boldsymbol{R}}$ in eq. \eqref{eq:molecular_fsd:molecular_matrix_element}, see \cite{PhysRevC.56.2132}. The transition operator resulting from the transformation from the lab to the molecular frame is only approximately correct in the form given in eq. \eqref{eq:molecular_fsd:molecular_matrix_element}.
The physical origin of the transition operator in the lab frame is that it describes transitions induced in the daughter molecule by the recoil of the departing $\upbeta$ electron and the neutrino. In view of the large mass difference and the fact that close to the end point of the $\upbeta$ spectrum basically all decay energy is given to the electron, the influence of the recoil induced by the neutrino is, however, negligible, see \cite{PhysRevC.56.2132}. The transformation to the transition operator in the molecular frame splits the effect of the recoil of the emitted electron into two parts. One is responsible for excitation of the center-of-mass motion of the molecular ion, {\it i.\,e.\ }for excitation of the translational degree of freedom, the other one for excitation of the internal molecular degrees of freedom (electronic, vibrational, and rotational). In the approximated form given in eq. \eqref{eq:molecular_fsd:molecular_matrix_element} only rotational and vibrational excitations can be induced, while the correction describing electronic excitation due to the recoil is ignored.
In \cite{PhysRevC.56.2132} the size of this effect, {\it i.\,e.\ }the probability for recoil-induced electronic transitions, was estimated based on a sum rule and found to be of the order of $5\cdot 10^{-5}$ and was neglected. 

Clearly, also the value of this fractional recoil $\tilde{K}=|\tilde{\bf K}|$ changes with the energy of the $\upbeta$-decay electron. This energy is reduced by the amount of energy left in the daughter molecular ion, so $\tilde{K}$ varies as a function of the energy at which the transition probability (and thus FSD) is evaluated.  This effect was, however, ignored in the KNM1-FSD, assuming it is negligibly small. An analysis of the uncertainty induced by this approximation is given in sec.\,\ref{constantfractionalrecoil}. 

Finally, an approximation is adopted with respect to the reduced mass used when solving for the rovibrational wavefunctions $\chi$ according to eq.~\eqref{eq:nuclear_eigenvalue_problem}. In principle, within the Born-Oppenheimer description, the Schr\"odinger equation describing nuclear motion should include the masses of the electrons, but \jannis{is} dependent on the electron density. To give a simplified example, consider if both electrons are in the case of $\HeT$ mostly at the He nucleus (ionic bond). In this case, both electron masses should be added to this nucleus\jannis{. If the electrons are equally distributed (covalent bond), then the mass of a single electron should be added to each nucleus.} Since the electron density depends evidently on the electronic state and even within one state it changes with internuclear separation, this effect is non-trivial to be included exactly into the calculation. Clearly, an $R$ dependent reduced mass (that needs to be obtained for every electronic state separately) should be included. In sec. \ref{reduced-mass} the most extreme case of inclusion and exclusion of the electron masses in the calculation of the reduced mass, but independent of the internuclear separation, is investigated. 
  
\subsubsection{Uncertainties on fundamental constants and conversion factors}\label{sec:conversion_factors}
The Hamiltonians used in the calculation contain fundamental constants like the electron mass, the Planck constant, etc. The Hamiltonian describing nuclear motion depends on the nuclear masses. 
As already discussed, the fractional recoil $\tilde{\bf K}$ depends on the energy of the $\upbeta$-decay electron, where the maximal electron energy (end-point energy $E_0$) is not exactly known. While the FSDs in \cite{Fackler1985} and \cite{PhysRevLett.84.242} used the value  $E_0=\SI{18.6}{\kilo\electronvolt}$, the KNM1 FSD adopted a more precise value of $\SI{18573.24}{\electronvolt}$ taken from \cite{dia:bodine2015}. 
Using different $E_0$ values in the FSDs results in an effect on the uncertainty similar to the neglect of the variation of $\tilde{K}$ within the 40\,eV interval discussed in sec.\,\ref{sec:deliberate_approximations}. This effect on $m_\nu^2$ is given in sec. \ref{endpointuncertainty}. 
     
In practice, all FSD calculations are performed using atomic units (Hartree), but for the analysis of KATRIN data the energies need to be converted into eV. This conversion factor depends on fundamental constants like the Planck constant or the elementary charge. The conversion factor changes over time as a result of more precise determinations of the fundamental constants. In fact, the corresponding recommended CODATA value changed in between the publication of the FSD by Fackler et al.~\cite{Fackler1985} and the calculation of the KNM1 FSD more than once. 
  
\subsubsection{Uncertainties due to convergence and precision}\label{sec:convergence_uncertainty}
For the case of the one-electron tritium atom and its decay to $^3$He$^+$, the transition probabilities can be calculated analytically if we assume to be non-relativistic, within the sudden approximation, and exclude electronic excitations due to recoil. The case of a two-electron atom requires on the other hand numerical solutions of the Schr\"odinger equation. In the case of a two-electron molecule like $\TT$, even within the Born-Oppenheimer approximation a numerical solution of the Schroedinger equation describing both electronic motion and nuclear motion is required, as was discussed in sec.\,\ref{sec:nuclear-motion}. An additional challenge is the need for a calculation of the energies and transition probabilities of all molecular final-states that lie within the energy interval used in the analysis of the $\upbeta$-decay spectrum, including possibly dissociation and ionisation continua. 
     
The use of explicitly correlated exponential basis functions in prolate-spheroidal coordinates known as Ko{\l}os-Wolniewicz functions, see eq.\,\eqref{eq:molecular_fsd:kolos_basis_functions}, for obtaining the solutions of the electronic Schr\"odinger equation of a diatomic two-electron molecule (and adopted in the Fackler FSD \cite{Fackler1985}, the Saenz FSD \cite{PhysRevLett.84.242}, and the KNM1 FSD \cite{valerian_paper}) has proven to be extremely efficient and accurate. It is, however, limited by the fact that one basis set is defined by one base, {\it i.\,e.\ }a single set of the non-integer parameters $\{\alpha, \bar{\alpha}, \beta, \bar{\beta}\}$ (Ko{\l}os-Wolniewicz) or $\{u,w,y,x\}$ (Pachucki {\it et al.}), see sec.\,\ref{omega-variation}. While a proper, judicious choice of the base allows for obtaining very accurate results for a given electronic state with a very small number of basis functions (\jannis{where} each basis function $j$ being defined by the integer quintuple $\{\mu_j,\lambda_j,\bar{\lambda}_j,\nu,\bar{\nu}_j\}$), formally an infinite number of basis functions would be required to describe an electronic state exactly. In the case of a finite basis set, the base should, in principle, be optimised individually both for every electronic state and internuclear separation R. However, the choice of the optimal base (non-integer parameters) depends also on the selection of basis functions (integer quintuples) and {\it vice versa}. 

Inputs from different calculations and basis sets have been used in the generation of the KNM1 FSD. For the electronic ground states of parent and daughter molecules highly optimised potential curves (including adiabatic corrections for all molecules, for the parent molecules T$_2$, HT, and DT also relativistic and radiative ones)\jannis{, see app. \ref{app:basis},} and electronic overlaps $S_n(R)$ were adopted from literature \jannis{\cite{dia:kolos85}}. Similarly, the potential curves and overlaps for the lowest lying five excited electronic states were taken from literature \jannis{(see \cite{dia:prc99})} where they had been optimised \jannis{individually } for the various states. The remaining \AS{(}Rydberg\AS{)} states \AS{(corresponding for the adopted basis to the electronic states 7 to 13)} were taken from literature where they had been obtained with a single basis set that was found to provide the best compromise for representing all excited states, as well as the electronic continuum. Clearly, while the accuracy of the electronic input data is supposed to be very high (and for the electronic ground-state potential curves this was confirmed by a comparison to various experimentally determined transition energies, see, {\it e.\,g.,} \cite{dia:doss_comparison,dia:prc99,dia:pach2012b}), the use of finite, rather than infinite, basis sets leads to an FSD uncertainty.
Note, it is easier to achieve the basis-set completeness \AS{and thus higher accuracy} in the one-dimensional equation of nuclear motion, eq.~\eqref{eq:nuclear_eigenvalue_problem}, compared to the six-dimensional electronic part. Of course, in all cases the question of finite precision arises, for example also in the quadrature used when calculating the transition probabilities, {\it i.\,e.\ }the integrals in eq. \eqref{eq:molecular_fsd:radial_integrals}. The effect of using a finite basis set is evaluated in sec. \ref{sec:potential_curves}, the effect of the choice of specific base parameters on $m_\nu^2$ in sec. \ref{sec:basis_excited}. The choice of ground-state energies, which is related to the convergence of the basis set, has an impact on $m_\nu^2$ as described in sec. \ref{sec:endpoint-reference}.

\subsubsection{Coding or input errors}\label{sec:code_input_errors}
Besides uncertainties there is\AS{, despite careful testing and cross-checking of the used codes and inputs,} evidently also the possibility of errors in the various computer codes used in order to evaluate the FSD and the quantities entering this calculation. Furthermore, input parameters may not be chosen properly, or typos may have occurred. Clearly, these are errors and not uncertainties, but in those cases where such errors were detected after the KNM1 FSD had already been used in the analysis of KATRIN data, it is evidently of interest to evaluate the effect that a given error has on the extracted $m_\nu^2$. 
Based on the analysis of the impact of a given error on the fit result it can be decided whether the experimental data need to be re-analysed with a properly corrected FSD, or whether the impact of the error is negligible with respect to the other uncertainties.

\subsection{FSD-uncertainty estimate adopted in KNM1\label{old-procedure}}
In the analysis of the first neutrino-mass measurement campaign of KATRIN the systematic contribution of the uncertainties of the FSD was estimated based on more general properties of the FSDs, namely, the moments of the distribution. A conservative estimate from a comparison of the \textit{ab initio} calculations by Saenz et al.~\cite{PhysRevLett.84.242} \AS{and the earlier ones by} Fackler et al.~\cite{Fackler1985} yielded for the electronic ground-state contribution to the FSD (energy interval below $\SI{10}{\electronvolt}$) an uncertainty of 1\,\% on the variance. However, \AS{in the generation of the Saenz FSD the same Born-Oppenheimer potential curves and corrections to them as well as the same electronic transition moments were used for the electronic ground states as were used in generating the Fackler FSD}, since the ones used in the latter were found to be already very accurate. Only for the electronically excited bound states slightly improved potential curves and matrix elements were used in the Saenz FSD in \cite{PhysRevLett.84.242}. The main difference between the two FSDs consisted in the treatment of the electronic continuum that, however, only contributes to the FSD at excitation energies above 40\,eV, i.e. to the $\beta$-spectrum below $E_0-40$\,eV.
It is thus not relevant for the first two science campaigns of KATRIN in which the fit range used in the analysis did only extend to 40\,eV below the end point. \AS{On the other hand, in \cite{dia:bodine2015} it was shown that the the non-physical $m_\nu^2$ in the results of the LANL~\cite{PhysRevLett.67.957} and LLNL~\cite{PhysRevLett.75.3237} experiments that were analysed with the Fackler FSD could be resolved, if the Saenz FSD with a better description of the continuum is adopted. 
Below 40\,eV the dominant difference between the Fackler FSD and the Saenz FSD is due to the use of non-relativistic and relativistic value for the fractional recoil $\tilde{K}$, respectively, as is explained in \cite{PhysRevC.60.034601}.} It leads to a shift of the mean excitation energy by approximately 0.03\,eV and a change in variance of the order of 1\,\%.\\
As a rather conservative uncertainty estimate for the FSD, and considering a good agreement of the theoretical predictions and the measurement of the probability of dissociated to non-dissociated molecular fragments by the TRIMS experiment\,\cite{TRIMS:2020nsv,TRIMS:unpublished}, a 1\,\% uncertainty on the normalisation of the ground to excited states populations was assumed. Based on these findings, the uncertainty of the total variance of the FSD was constrained to 2\,\%. The corresponding uncertainty of the variance of the electronically excited states and ionisation continuum was set to 4\,\%.

The FSD uncertainty estimated this way was propagated into the $m_\nu^2$ uncertainty using \AL{two different techniques: the covariance matrix and the Monte-Carlo propagation. In both approaches the } binned FSD was modified randomly in a bin-to-bin uncorrelated way to provide the corresponding variations of the electronic ground-state variance (1\,\%), its probability (1\,\%), and the variance of the electronically excited states (4\,\%). Repeating this modification many times and calculating the corresponding integrated spectra a covariance matrix was built for the FSD-related uncertainty. \AL{The constructed covariance matrix in the first approach and randomised FSDs in the Monte-Carlo propagation method were used in the spectral fit of the data.} The corresponding variation of the $m_\nu^2$ parameter yielded the estimate of the additional uncertainty. In a typical narrow analysis interval of 40\,eV below the end point $E_0$ the electronic ground state contributes the largest fraction to the measured spectrum and the FSD-related $m{_\nu}^2$ uncertainty was given as $\SI{0.02}{\square\electronvolt\per c^4}$ \cite{KNM1Paper,AnalysisPaper}. 
  
For the first two science campaigns of KATRIN this uncertainty estimate for the FSD was sufficient, since even the very conservative estimate indicated that the FSD contribution to the uncertainty budget was sufficiently small compared to the statistical uncertainty. However, with the increase of the statistics and other systematic uncertainties being reduced, it is crucial to provide a more stringent way to determine the FSD uncertainty; also because an unjustified too conservative estimate would unnecessarily limit the capability of KATRIN. This new way of uncertainty determination should also discriminate the contribution of each of the types of FSD uncertainties discussed in sec. \ref{types}.

\section{New FSD uncertainty analysis}\label{sec:new_uncertainty}
In this chapter, the new procedure for determining the FSD uncertainty via a basis-set convergence approach is introduced. The general requirements for the uncertainty analysis are motivated and described in sec. \ref{sec:general-estimation}. As will be shown in sec. \ref{sec:pseudo_KNM1}, the new way of uncertainty determination cannot be applied to the KNM1 FSD as is. 
Finally, the concrete procedure of the uncertainty determination which fulfills the requirements listed in sec. \ref{sec:general-estimation} is presented in sec. \ref{sec:5.3}.
\subsection{General strategy for the uncertainty estimation\label{sec:general-estimation}}
As discussed in sec. \ref{old-procedure}, an estimate of the FSD uncertainty based on the agreement of the Fackler and the Saenz FSDs \AS{as was done previously for the uncertainty estimate of the KNM1 FSD} may be not robust enough: there is a substantial overlap both of the inputs and the computational approaches. This includes the potential curves and electronic overlaps for the electronic ground states of parent and daughter molecules. Although improved data were used for the electronically excited states, they were still calculated with the same type of basis functions, Ko{\l}os-Wolniewicz geminals. The Saenz FSD contains a substantial improvement in the electronic continuum, but this part of the FSD is only relevant, if the fit range of an experiment extends beyond 40\,eV \jannisfourty{below the end point}.
  
While there exists no direct experimental measurement of the molecular FSD besides the already mentioned recent re-determination of the dissociation branching ratio by the TRIMS experiment \cite{TRIMS:2020nsv,TRIMS:unpublished}, there \jannis{are} numerous reasons and data that provide confidence in the theoretically obtained FSDs (Fackler FSD, Saenz FSD, and KNM1 FSD). The potential curves of the electronic ground states of both the parent and the daughter molecules have been evaluated \jannis{multiple times with various approaches, \jannis{see {\it e.g.} \cite{PhysRevLett.99.240402,RYCHLEWSKI1994657,dia:pachucki_relativistic_h2,CENCEK1995417,dia:pach2012}.}} Furthermore, there exists a wealth of spectroscopic data in which the transition energies between many rovibrational states have been very accurately measured for a number of isotopologues, \cite{dia:pach2012b}. In \cite{PhysRevC.60.034601} the energies obtained when adopting the potential curves that were used in generating the Fackler FSD were compared with numerous spectroscopic data and extremely good agreement was found. This finding together with the fact that basis-set variations, though also using Ko{\l}os-Wolniewicz geminals, did not lead to lower energies, and thus not to better results (as can be uniquely concluded on the basis of the variational principle), motivated their use in the evaluation of the Saenz FSD. Noteworthy, also the improvement of the \AS{bound} electronically excited states \AS{only} led to very small changes of the FSD. In view of the requirements of KATRIN (higher temperature \jannis{and the presence of the} additional isotopologue DT), a new FSD was \jannis{calculated} in \cite{PhysRevC.73.025502,dia:doss_thesis}, but using the same input and codes as used in the evaluation of the Saenz FSD. Furthermore, a comparison to new spectroscopic data was performed, again finding very good agreement \cite{dia:doss_thesis}.  Also more recent improved calculations of the electronic ground states of the parent and daughter molecules \cite{dia:pach2010,dia:pach2012} only increased the number of digits, but did not modify the potential curves adopted in the Fackler, Saenz, and KNM1 FSDs within the precision (number of digits) given therein.
  
There is much less confirmation of the correctness of the potential curves of the electronically excited states. In the case of the electronic transition matrix elements (always involving the electronic ground state of the parent molecule) there could even have been implementation errors, cf.\ sec.~\ref{sec:code_input_errors}. Therefore, before the KNM1 FSD was generated, numerous checks were performed that will be reported elsewhere \cite{checks_to_be_published}. For example, the electronic problem (potential curves, wavefunctions, and transition matrix elements) was additionally solved adopting a different method (configuration interaction adopting a $B$-spline basis in prolate-spheroidal coordinates \cite{dia:vanne}) and thus also using completely independently developed computer codes. The convergence behaviour was much slower compared to the geminal approach. Nevertheless, within the convergence that could be achieved, the input data (potential curves and matrix elements) used for the Fackler and Saenz FSDs were validated. Adopting Ko{\l}os-Wolniewicz geminals, but the new H2SOLV code \cite{CPC} that allows for systematic basis-set improvements, again the high accuracy of the previously used FSD input data was confirmed. 
  
Also the nuclear problem that was solved before by adopting numerical integration based on the Numerov method was re-evaluated with a different approach (expansion of the radial part in $B$-splines) and thus again a completely independently written computer code. Also in this case very good agreement was found. All these findings motivated the way the KNM1 FSD was generated and used in the analysis of the first two science campaigns of KATRIN.
  
Though the numerous convergence studies provided validation of the generated FSD, they do not provide an explicit uncertainty, i.\,e.\ a single value specifying the contribution of the uncertainty of the FSD to the overall systematic uncertainty of $m_{\nu}^2$ extracted from the KATRIN data. For example, the accuracy of the adopted potential curves as estimated from the convergence studies depends on the electronic state, and for a given state on the internuclear separation. An uncertainty in the potential curve leads to an uncertainty in the energies of the rovibronic \jannis{(rotational, vibrational and electronic)} states, \jannis{simultaneously} the electronic wavefunction (the eigenvector belonging to a given potential curve and thus the eigenvalue) and its possible uncertainty influences the electronic transition matrix elements. Clearly, the uncertainties are highly correlated in a non-trivial way. 
  
\AS{In contrast to the situation for the electronic ground state, the accuracy of the electronically excited states and their uncertainties are much less precisely known due to the lack of alternative calculations, but especially due to the lack of experimentally determined spectroscopic data}. However, the influence on the extracted $m_\nu^2$ is much larger for the ground state, but its relative importance depends on the fit range included in the analysis.    

There are thus different aspects that should be considered in the new uncertainty analysis. First of all, it has to depend on the experimental conditions\jannis{:} the energy interval included in the fit determining the molecular final states that contribute to the FSD, the temperature and the isotopologue distributions \jannis{which} affect the statistical \jannis{weights} with which different initial states contribute, the energy resolution requiring some minimal bin sizes, etc. Second, the impact on the extracted $m_\nu^2$ by the individual sources of uncertainties should become transparent, also allowing for the identification of those ingredients or approximations entering the FSD that need to be improved most urgently, if the FSD uncertainty needs to be reduced. \jannis{In this way, artificial reduction of uncertainties due to counteracting or canceling effects may be identified.} Third, the accuracy of the potential curves, transition matrix elements, and corrections that are calculated with \AS{finite-basis-set approaches} should be assessed based on systematic basis-set convergence studies. The uncertainty analysis should thus incorporate correlations between, e.\,g., the quality of the potential curves and the one of the corresponding wavefunctions, in a consistent way. Most importantly, the variable accuracy of the input data, {\it e.\,g.\ }the higher accuracy for the electronic ground states compared to the electronically excited ones, as well as their impact depending on the fit interval should be contained in the uncertainty analysis. The third aspect, \AS{i.\,e.\ the systematic basis-set convergence study that, however, cannot be performed with the KNM1 FSD itself due to the mixed bases and partly missing input information,} motivates the need for a pseudo-KNM1 FSD for the studies presented here. In the following sec. \ref{sec:pseudo_KNM1} the generation of this pseudo-KNM1 FSD is described, while sec.\,\ref{sec:5.3} introduces the new procedure and discusses how these requirements are fulfilled.

\subsection{Need of a pseudo-KNM1 FSD}\label{sec:pseudo_KNM1}
Applying the proposed new uncertainty analysis to the KNM1 FSD results in some difficulties. First of all, different basis-sets (different values of both the base $\{\alpha,\bar{\alpha},\beta,\bar{\beta}\}$ and basis-set expansions, {\it i.\,e.\ }the number and the selection of the quintuples $\{\mu_j,\lambda_j,\bar{\lambda}_j,\nu_j,\bar{\nu}_j\}$, see eq. \eqref{eq:molecular_fsd:kolos_basis_functions} in sec.\,\ref{sec:potentialcurves}) had been adopted in the evaluation of the electronic input data. Even worse, not the complete information of all involved basis sets at all internuclear separations entering the previous calculations could be fully recovered. For example, there is some cross-referencing referring finally to a reference that does not contain a sufficiently detailed information. Sometimes, simple typos are evident (the same basis function appearing twice) but nevertheless not (simply) recoverable (as it is unclear which basis function was used instead) or there were corrections given in some other publication, but it was not found to be certain whether this correction is correct itself. In fact, in some cases the canonical orthogonalisation had been adopted to handle numerically induced linear dependencies. Even if the cut-off threshold used for reducing the basis set is given, the results depend on the order of the basis functions (and thus the quintuples), but this information is not always available. Even the numerical precision of the adopted hardware and compiler, but especially the adopted expansion length in the von Neumann expansion used for solving the integrals may influence the result of the orthogonalisaation and this information is often missing. It should be emphasised that these problems apply only to some of the input values. 
For all internuclear separations that contribute substantially to the FSD the achieved agreement to the literature values is $10^{-6}$ to $10^{-9}$. Most of the input values could be reproduced to a sufficient degree (after a corresponding very laborious trial-and-error procedure based on inverse engineering), see app.~\ref{app:FSD_parameters} for details. Also, the input values itself (potential-curve values including the corrections to them or electronic overlaps) that entered the KNM1 FSD are all reproducible in the sense that every value used and taken from literature can be found in the correspondingly cited publications. It was thus concluded that the demands in computational resources and the time that would be needed to recover every single basis-set parameter for every input value are too large while the impact on the here performed uncertainty analysis would be completely negligible in order to justify the attempt of a full input-data recovery. It should be again emphasised that these data are solely required for a fully consistent uncertainty analysis as it is proposed here, but not for the reproduction of the KNM1 FSD itself. Due to the difficulties described above, it was decided to create a pseudo-KNM1 FSD, which is as close as possible to the KNM1 FSD, but for which all input parameters are known and can thus be systematically varied. The generation of the pseudo-KNM1 FSD is based on the input data and basically the same procedure and numerical apparatus used for obtaining the KNM1 FSD \cite{valerian_paper}. Most importantly, the pseudo-KNM1 FSD allows for the systematic basis-set enlargement required for the uncertainty analysis. 
Clearly, the need for a pseudo FSD stems from the way the KNM1 FSD was constructed that is not suitable for a systematic and consistent uncertainty analysis. This motivated the use of a differently obtained FSD for the analysis of more recent KATRIN measurement campaigns \cite{KNM5Neutrino,FSDKNM5}. 

\subsection{Procedure to estimate the KNM1 uncertainty\label{sec:5.3}}
In view of the requirements the uncertainty analysis should fulfill, the following procedure is proposed and is illustrated by its application to the KNM1 FSD adopting the experimental conditions from the KNM1 measurement campaign. With the aid of a theoretically obtained pseudo-KNM1 FSD a $\upbeta$-decay spectrum is generated on the basis of the experimental parameters, setting the neutrino mass to zero. The pseudo-KNM1 FSD is close to the KNM1 FSD, but differs from the the latter in the following ways:

\begin{figure}
  \begin{subfigure}{.5\textwidth}
    \centering
    \includegraphics[height=5cm, page = 1]{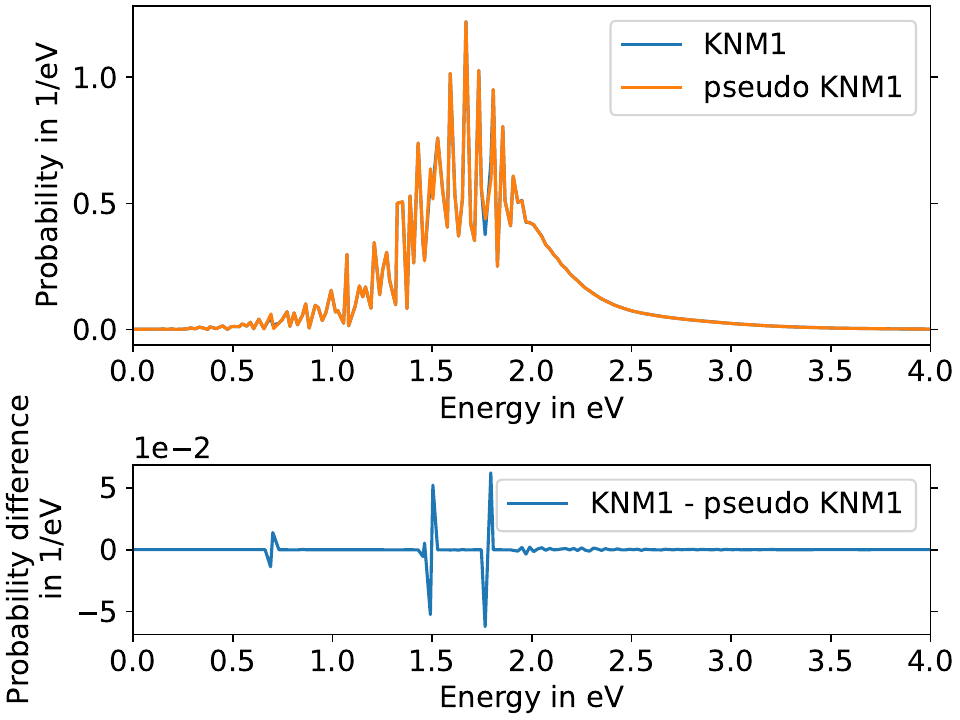}
  \end{subfigure}
  \begin{subfigure}{.5\textwidth}
    \centering
    \includegraphics[height=5cm, page = 2]{knm1_knm1_ref_comparison.pdf}
  \end{subfigure}
  \caption{
    FSD for $\TT$ up to \SI{40}{\electronvolt} (corresponding to \SI{40}{\electronvolt} below the end point) as used in the analysis of the first KATRIN science campaign (KNM1 FSD), in comparison to the FSD for $\TT$ used for the uncertainty investigations in this work (pseudo-KNM1 FSD). \jannis{On the left hand side the ground state is shown, while on the right hand side the excited states can be seen.}
  }
  \label{fig:introduction:}
\end{figure}

\begin{itemize}
    \item The pseudo-KNM1 FSD uses nuclear reduced masses $\mu_\text{n}$ for the parent nuclei of all isotopologues and effective reduced masses $\mu_\text{eff}$ for the daughter nuclei, while the KNM1 FSD adopts simultaneously different types of masses, {\it i.\,e.\ }not all daughter isotopologues use $\mu_\text{eff}$ (see sec. \ref{reduced-mass}) 
    \item The pseudo-KNM1 FSD uses a density approach \cite{valerian_paper} with probability densities for the description of the dissociation continua of the electronically excited states, while the KNM1 FSD uses a discretised approach. The consequences of using either of the approaches is discussed in app. \ref{app:dissociation}.
    \item The pseudo-KNM1 FSD is constructed with a basis-set convergence parameter $\Omega=10$, while the basis set of the KNM1 FSD corresponds effectively to values between $\Omega=5$ and $\Omega=7$ (see sec. \ref{omega-variation})
\end{itemize}
While the Born-Oppenheimer potential curves are independent of the nuclear masses, this is not the case\jannis{ in the adiabatic approximation. In the adiabatic case} individual potential curves for the different isotopologues T$_\text{2}$, $\DT$, and $\HT$ as well as the corresponding daughter isotopologues are used in the FSD calculation.  
A comparison of the KNM1 FSD and the pseudo-KNM1 FSD is shown in fig. \ref{fig:introduction:}. The impact of the difference between adopting either the KNM1 FSD or the pseudo-KNM1 FSD on $m_\nu^2$ is discussed in app. \ref{sec:fsd-comparison}.

In order to investigate a specific uncertainty, the $\upbeta$-decay spectrum generated using the pseudo-KNM1 FSD  is fitted by a $\upbeta$-decay spectrum using a {\it test FSD} (or set of test FSDs) describing this specific uncertainty (see eq. \eqref{eq:introduction:beta_decay_fermi}) and the $m_\nu^2$ value is extracted.
The resulting shift in $m_{\nu}^2$ provides a direct measure of the impact of the specific uncertainty. This specific uncertainty could be 
a specific limitation of the calculation (like the finite basis set), an adopted approximation (like the sudden approximation), or some input error.
  
Clearly, the result depends on the basis set adopted in the evaluation of the Born-Oppenheimer potential curves and electronic transition matrix elements. Therefore the investigation is repeated with a set of test FSDs with systematically improved basis sets. The values obtained for $m_\nu^2$ by a fit on the Monte-Carlo data set described in sec.~\ref{generation} in which always the same pseudo-KNM1 FSD is used should then systematically converge to $m_\nu^2=0\,$\,eV$^2$/$c^4$.
Its deviation from this value provides a quantitative estimate of the uncertainty due to the considered limitation, approximation, or input error independently of the basis-set limitation. The results on the individual FSD uncertainties obtained by using the procedure described in this section will be discussed in sec. \ref{sec:results}.

%--------------------------------------------------------------------------------
% Results and discussion
%--------------------------------------------------------------------------------
\section{Results and discussion}\label{sec:results}

In this chapter, the  results for the individual uncertainty contributions from the FSD calculation onto $m_\nu^2$ are presented (see sec. \ref{sec:individual-contributions}). To achieve consistency in the estimation of the uncertainty, all of the following plots use the same Asimov Monte-Carlo data set generated with the pseudo-KNM1 FSD specified in sec. \ref{sec:new_uncertainty} and $m^2_\nu=0$ as input. The other parameters of the data set are as described in sec. \ref{run-condition} and sec. \ref{generation}.
The KNM1 FSD and the pseudo-KNM1 FSD are similar though their differences do introduce
a tiny and constant shift in all fit results for $m_\nu^2$, see app. \ref{sec:fsd-comparison}. Besides the constant shift, the FSD used for the Monte-Carlo data does not impact the results from the convergence studies.

Unless otherwise stated, the graphs in this section all comprise the same $y$ axis. They show the fit result for $m_\nu^2$ when fitting the same Asimov Monte-Carlo data set described above while using anotherthan the pseudo-KNM1 FSD for the fit model which differs from the pseudo-KNM1 FSD by one parameter of interest. With two exceptions, these fit values are plotted against $\Omega$ of the basis set (see sec.~\ref{omega-variation}) on the $x$ axis. The FSDs generated with the convergence parameter set to $\Omega=10$ appear to be a good compromise between convergence that is achieved and computational efforts. Therefore, the deviation of the $m_\nu^2$ obtained for $\Omega=10$ gives the uncertainty associated with the effect under investigation.\footnote{In principle, this holds only for a fully converged FSD. Since the convergence study can, however, for the reasons discussed not be performed for the KNM1 FSD, the present assignment of the individual uncertainties is chosen. If the $m_\nu^2$ uncertainties obtained from the convergence studies were evaluated at some approximate value of $\Omega$ reflecting the KNM1 FSD, the effect of a given source of uncertainty would be entangled with the calculation of the uncertainty itself, since the correction was calculated using a not-converged basis set. The same uncertainty would thus be considered multiple times.} In the following sec. \ref{sec:individual-contributions}, all individual uncertainty contributions which are obtained following the described procedure are listed. The final new total uncertainty for the KNM1 FSD which is obtained is presented in sec. \ref{sec:final-uncertainty}.

\subsection{Individual uncertainty contributions\label{sec:individual-contributions}}
In the following, the individual effects leading to an FSD uncertainty are discussed separately. They are the impact of the convergence parameter $\Omega$ (see sec. \ref{sec:potential_curves}), the choice of the base for \jannis{describing} the excited states (see sec. \ref{sec:basis_excited}), the impact of the inclusion of theoretical corrections (relativistic, radiative, and adiabatic corrections, see sec. \ref{corrections}), the choice of the reduced masses (nuclear or effective, see sec. \ref{reduced-mass}), the influence of the molecular ground-state energies (see sec. \ref{sec:endpoint-reference}), the influence of binning (see sec. \ref{subsec:binning}), the impact of the uncertainty on the tritium Q value (see sec. \ref{endpointuncertainty}), the neglect of the fractional-recoil variation with energy (see sec. \ref{constantfractionalrecoil}) and the use of the sudden approximation (see sec. \ref{subsec:suddenapprox}).

%--------------------------------------------------------------------------------
% Omega convergence
%--------------------------------------------------------------------------------
\subsubsection{$\Omega$ convergence}\label{sec:potential_curves}

The \jannis{electronic part of the} original KNM1 FSD was computed using a variant of a code originally written by Ko{\l}os and collaborators (see sec.~\ref{sec:potentialcurves}). The basis sets used in the KNM1-FSD calculation entering eq.~\eqref{eq:molecular_fsd:kolos_basis_functions} had been taken from literature, \jannis{as is described in app. \ref{app:basis}}. Therein, they had been optimised individually as a function of the internuclear separation for the ground electronic states of $\HH$ and HeH$^+$ with respect to both the base (non-linear parameters) and the basis functions (quintuples) \cite{valerian_paper}. For the electronically excited states of HeH$^+$ 400 basis functions (integer quintuples) and three different sets of the base parameters $\alpha,\; \bar{\alpha},\; \beta\; \mathrm{and}\; \bar{\beta}$ were chosen. \AS{Out of the results obtained with the different bases, for every one of the five lowest lying electronically excited states and every value of the internuclear separation the lowest energies (and corresponding overlaps) were} chosen. For some internuclear separations and excited states the potential curves and transition matrix elements already used in the Fackler FSD \cite{Fackler1985} were adopted, if the corresponding energies were lower\AS{. A}ccording to the variational principle both energies and wavefunctions are \AS{then} more accurate. Evidently, \AS{such a combination of individually selected data obtained with different basis sets} cannot be mapped onto a simple $\Omega$ variation. However, such a mapping would be required for a clear systematic uncertainty analysis as is proposed in this work. \jannis{Therefore for the KNM1 FSD the result is approximate.} In order to validate the approximation, some consistency checks are performed. 

A first assessment of the quality of the KNM1 FSD is obtained by a comparison of the Born-Oppenheimer potential curves $V_{n}^{\rm BO}$ from KNM1 with those obtained in the present work for the different values of $\Omega$, both for the ground states of $\HH$ and of HeH$^+$. Since the ground-state population of $\HeH$ after $\upbeta$ decay of $\TT$ is about 57\% of the total probability \cite{TRIMS:2020nsv}, it is most relevant and thus chosen. From such a comparison (not shown) it is concluded that the potential curves used in generating the KNM1 FSD are of a quality similar to the ones obtained with $\Omega = 5 - 6$ for HeH$^+$ and $\Omega = 6 - 7$ for $\HH$. Therefore, it is to be expected that the results of the KNM1 FSD should be comparable to test FSDs obtained with values of $\Omega = 5 - 7$. 

\begin{figure}[htpb]
    \centering
    \includegraphics[width=0.60\textwidth, page = 3]{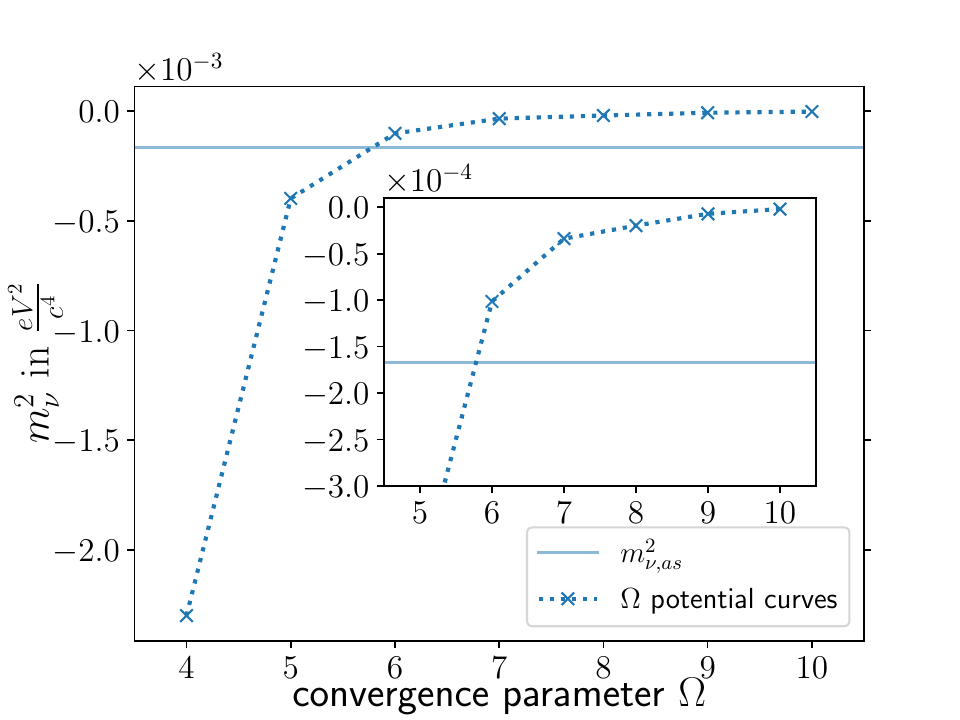}
    \caption{
     Convergence curve using $\Omega$-dependent potential curves in comparison to the horizontal $m_{\nu, \text{as}}^2$ line obtained when the original potential curves from the KNM1 FSD are used.  
    }
    \label{fig:results:omega_convergence}
\end{figure}

In order to provide an estimate of the KNM1-FSD uncertainty with respect to the individual uncertainty sources, the KNM1 FSD potential curves are associated an $m_{\nu, \text{as}}^2$ that is obtained from using a test FSD with the KNM1 potential curves in the fit model for the pseudo-KNM1 Monte-Carlo data set. This way it can be determined where the single-valued KNM1-FSD uncertainty is placed on the $\Omega$ convergence curve. This allows for the assignment of an effective value of $\Omega_\text{eff, KNM1}$. This is illustrated in fig. \ref{fig:results:omega_convergence} where the horizontal line shown is the aforementioned value $m_{\nu, \text{as}}^2$ \jannis{corresponding to the extracted $m_\nu^2$ when KNM1-FSD potential curves are used}, while the crosses (connected by a dotted line to guide the eye) show the convergence curve which shows the results obtained for an increasing value of the convergence parameter $\Omega$. 
The intersection of the two curves delivers the estimate $\Omega_\text{eff, KNM1}\approx 6$. The difference between $m_{\nu, \text{as}}^2$ and the $m_\nu^2$ of the convergence curve at $\Omega = 10$ delivers the error of the KNM1 FSD compared to an FSD that is obtained for potential curves that are supposed to be converged. The error found this way is \SI{-1.7e-4}{\square\electronvolt\per c^4}.\\ 

%--------------------------------------------------------------------------------
% basis parameters excited states
%--------------------------------------------------------------------------------
\subsubsection{Base parameters}\label{sec:basis_excited}
As discussed earlier, each of the potential curves produced with the H2SOLVm code depends only on the two inputs specifying the basis set: the base, {\it i.\,e.\ }$\{\alpha$, $\bar{\alpha}$, $\beta$, $\bar{\beta}\}$, and the convergence parameter $\Omega$ (see sec. \ref{omega-variation}).
As the KNM1 FSD adopts the potential and overlap curves from literature, the base $\{\alpha$, $\bar{\alpha}$, $\beta$, $\bar{\beta}\}$ used to produce these curves is needed in order to perform a convergence study. However, complete information on a base (that usually varies with the nuclear separation, and for different electronic states) wasn't always available in the references. In these cases, an attempt was made to reconstruct the bases. Additionally the choice of base is limited by the fact that the recurrence relation used for solving the integrals in H2SOLV imposes certain restrictions on the bases that can be used. This excludes some of the bases used in generating the KNM1 FSD (where the Ko{\l}os code was adopted). All bases that were used, but could not be found in literature, are given explicitly in app. \ref{app:basis}. The bases adopted for the ground states, even the ones for $\HeH$ that were obtained by a reconstruction, yield very accurate results (low energies) as they had been very carefully optimised. Thus only the much larger influence of the choice of different bases on the excited states is discussed in this work.

\begin{figure}[htpb]
    \centering
    \includegraphics[width=0.60\textwidth, page = 4]{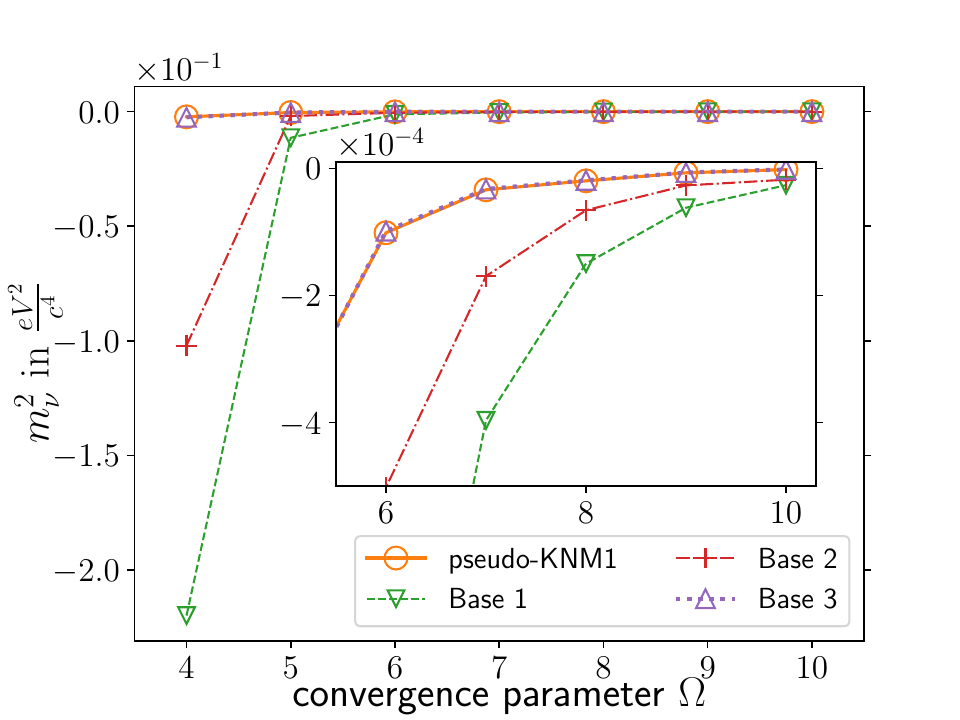}
    \caption{
        Comparison of the convergence behaviour of the extracted squared neutrino mass for different bases used for the computation of the electronically excited $\HeH$ states. 
        The results obtained for the three different Bases 1, 2, and 3 are compared with the results of the 
        pseudo-KNM1 FSD. Those three bases (see text and app. \ref{basis} for details) are constituents of the combined base in the pseudo-KNM1 FSD calculation.}
    
    \label{fig:results:bases_shift}
\end{figure}

The potential curves of the excited states in the KNM1 FSD are composed of four different bases, as described in app. \ref{basis}. For each of the excited states $n$ the lowest energy for a given internuclear separation $R$ was chosen for constructing the corresponding potential curve $V_n^{\rm BO}(R)$ and electronic overlap matrix element $S_n(R)$. 
In order to estimate the uncertainty due to the choice of the base, for each of the three different (known) bases that were used in generating the pseudo-KNM1 FSD a separate FSD was calculated as a function of the convergence parameter $\Omega$ (based on the corresponding base-dependent potential curves and electronic overlaps for the electronically excited states of $\HeH$). The convergence of the squared neutrino mass  $m_{\nu}^2$ obtained from using these three different sets of FSDs is shown in fig.~\ref{fig:results:bases_shift} and compared with the results for the pseudo-KNM1. 

Despite the fact that the different bases show quite a different convergence behaviour, all curves converge for larger values of $\Omega$ to $m_{\nu}^2$ values which are in agreement with each other to within \SI{3e-5}{\square\electronvolt\per c^4} for $\Omega=10$. The results in fig.~\ref{fig:results:bases_shift}
confirm the expectation that a poor choice of the base (large deviation from an accurate calculation) for small values of $\Omega$ can be compensated by using a large value of $\Omega$, since for sufficiently large values of $\Omega$ the results for different bases converge to the same value.  

%--------------------------------------------------------------------------------
% corrections to born oppenheimer
%--------------------------------------------------------------------------------
\subsubsection{Relativistic, radiative, and adiabatic corrections\label{corrections}}
The potential curves obtained within the non-relativistic Born-Oppenheimer approximation can be improved by taking various corrections perturbatively into account (see sec. \ref{sec:deliberate_approximations} for details). In the KNM1 FSD such corrections were only adopted for the electronic ground states of the parent and daughter molecules. Since the initial states of the parent molecule enter the calculation of all transition matrix elements and more than half of the total transition probability goes into the electronic ground state of the daughter molecule, it is expected that corrections are most important for these states. \AS{Since the largest corrections are available in literature and are assumed to be sufficiently accurate, they were adopted.} \jannis{If these adopted corrections are only available for a finite set of internuclear separations, they were taken as is, spline interpolated, and applied to the corresponding potential curves.} This procedure offers the flexibility to add the corrections to any of the potential curves generated for creating test FSDs, even if they are not available for the same values of the internuclear separation $R$.

The most important correction to the non-relativistic Born-Oppenheimer potential curve is the already mentioned adiabatic (or diagonal) correction, {\it i.\,e.\ }the matrix element of the nuclear kinetic-energy operator between the same electronic wavefunctions in bra and ket. Since the adiabatic correction originates from the nuclear motion, it is mass dependent and has to be rescaled for the different isotopologues accordingly. For the parent isotopologues the values from \cite{dia:woln93} were taken. For the daughter isotopologues the data were taken from \cite{dia:bish79}. The correction from \cite{dia:bish79} was interpolated with a polynomial, as proposed in that work. For distances $R > 6\, a_0$ the value was assumed to be identical to that at $R=\infty$. The reason for this simplified treatment is twofold. First, the $R$ dependence of the correction at large internuclear separations is small, since the molecular wavefunctions become almost identical to the ones of two independent atoms and thus approximately $R$ independent. Second, due to the Franck-Condon principle the transition probability (in the present case the FSD) depends only on the $R$ interval in which the intitial-state wavefunction (here the one of the electronic ground state of the parent molecule) has a non-negligible density. The non-adiabatic corrections that are the off-diagonal matrix elements of the kinetic-energy operator of the nuclei are neglected as is explained in sec. \ref{sec:deliberate_approximations}. 

The (leading) relativistic correction is to a good approximation nuclear-mass independent as it only describes the effect of the spin and the velocity of the electrons, since they are much faster. The smallest of the corrections adopted here is the radiative correction, describing the interaction of the electrons with the quantized electromagnetic vacuum field. Both the relativistic and radiative corrections are taken from \cite{dia:woln93}, where they are provided for the parent isotopologues. To our knowledge there are no relativistic and radiative corrections in literature for the daughter isotoplogues. Therefore, and because their effect for the parent molecule is found to be very small (see below), they are neglected in the present uncertainty study.

Since the corrections discussed in this subsection were not re-calculated in this work, they are $\Omega$ independent. The adiabatic and relativistic corrections of \cite{dia:woln93} can be compared with the more recent and more accurate adiabatic corrections in \cite{dia:pachucki_adiabatic_h2} as well as the relativistic corrections in \cite{dia:pachucki_relativistic_h2}. The {\it relative error} between the adiabatic corrections used in the KNM1 FSD and given in \cite{dia:pachucki_adiabatic_h2} is at most of the order of $10^{-4}$, the absolute magnitude of the correction itself being only $\approx 5.5 \cdot 10^{-4}$. For the relativistic corrections used in the KNM1 FSD and given in \cite{dia:pachucki_relativistic_h2} the relative error is at most of the order of $10^{-3}$, the absolute magnitude of the correction itself being $\approx 1.5 \cdot 10^{-5}$. Therefore, the accuracy of those corrections was clearly sufficient for the KNM1 measurement campaign \AS{and they were adopted in both the KNM1 FSD and the pseudo KNM1-FSD calculation.}

\begin{figure}[htpb]
    \centering
    \includegraphics[width=0.60\textwidth, page = 6]{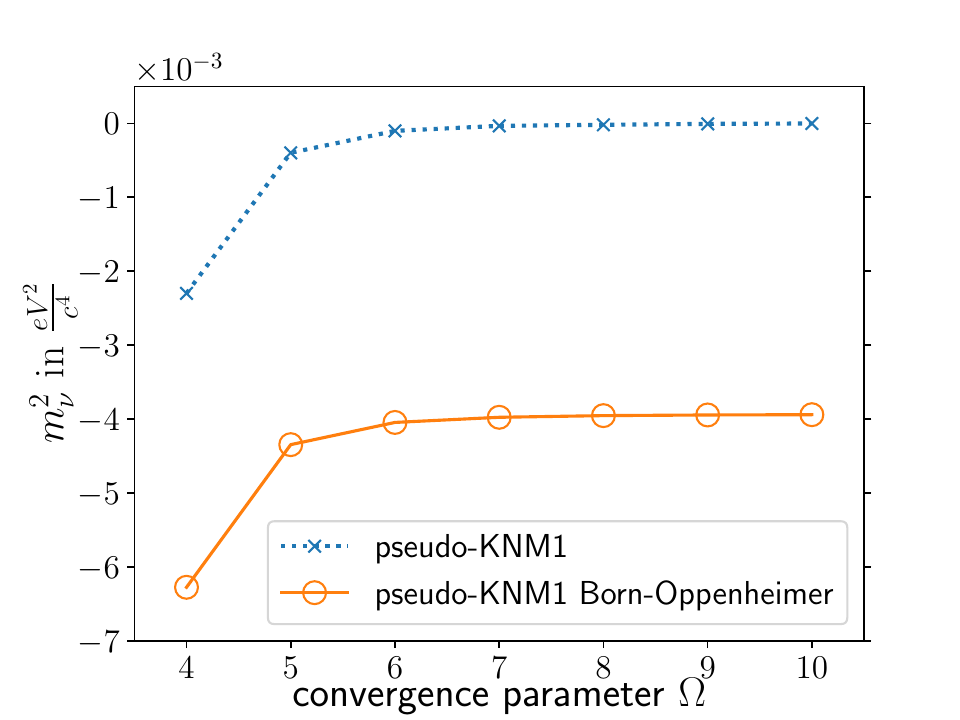}
    \caption{Comparison of the extracted squared neutrino mass obtained if corrections to the Born-Oppenheimer potential curves are applied or not. For the orange curve (label "pseudo-KNM1 Born-Oppenheimer"), the non-relativistic Born-Oppenheimer potential curves are used for the electronic ground states of parent and daughter molecules in the FSD calculation. For the blue curve (label "pseudo-KNM1"), the adiabatic, relativistic, and radiative corrections adopted in the KNM1 FSD are included. }
    \label{fig:results:corrections_shift}
\end{figure}

In order to estimate the uncertainty on $m_\nu^2$ due to the adiabatic, relativistic, and radiative corrections and the limited accuracy with which they were calculated, the extreme cases are considered in which the corrections are either \jannis{ all included or excluded.} The results, given as a function of the basis-set convergence parameter $\Omega$, are shown in fig.~\ref{fig:results:corrections_shift}. The complete omission of the corrections yields a constant shift of \SI{-4e-3}{\square\electronvolt\per c^4} for the extracted $m_\nu^2$ independent of $\Omega$. As a very conservative estimate for the uncertainty of $m_\nu^2$ due to \AS{the omission of the corrections for the electronically excited states} (and the limited accuracy of the adopted literature values for those corrections), 30\,\% of the absolute value of the uncertainty \AS{obtained when completely omitting these corrections for the electronic ground state} is used in the total uncertainty budget, since within the fit interval roughly 30\,\% of the FSD goes into the electronically excited states.\footnote{\AS{About 57\,\% of the total FSD goes into the electronic ground state and about 17\,\% lies \jannisfourty{\includecomment{outside} below} the 40\,eV fit interval.}}

%--------------------------------------------------------------------------------
% reduced mass
%--------------------------------------------------------------------------------
\subsubsection{Reduced mass}\label{reduced-mass}
When solving for the wavefunction of nuclear motion one needs to use the reduced mass $\mu$ of the parent and of the daughter molecules. This raises the question how to choose a proper value for the reduced mass $\mu$ (see end of sec.~\ref{sec:deliberate_approximations}). In \cite{dia:doss_thesis} four different ways are discussed for choosing the reduced mass (in the case of the electronic ground state) and the results are compared to spectroscopic data. In the present work, two of these options are considered: the nuclear reduced mass $\mu_\text{n}$ which only takes into account the masses of the nuclei without the electrons and the effective reduced mass $\mu_\text{eff}$ that distributes the masses of the two electrons according to the electron density. In the latter case it is assumed that one electron is bound to the $\text{He}^{2+}$ nucleus and the other one is distributed between both nuclei. The effective reduced mass is also chosen, because in \cite{dia:doss_thesis} it is demonstrated that using $\mu_\text{eff}$ gives better agreement with spectroscopy data, if used for the daughter molecules $\HeH$ and $\HeD$. For both the effective reduced mass and the nuclear reduced mass, the same relation 
\begin{equation}
    \mu = \frac{m_1 m_2}{m_1 + m_2}
    \label{eq:molecular_fsd:reduced_mass}
\end{equation}        
applies, but different values are used for the masses \AS{$m_1,m_2$, either ignoring or including the electron mass}. The nuclear reduced mass and effective reduced mass thus represent somehow the most extreme cases out of the four cases considered in \cite{dia:doss_thesis}.

\begin{table}[htpb]
    \centering
    \caption{
        The values of the nuclear and effective reduced masses (given in terms of the electron mass $m_{\text{e}}$) used in this work for the uncertainty estimation of the FSD are given together with  the values used in the calculation of the KNM1 FSD.}
    \begin{tabular}{llll}
        \toprule
                          & T$_2$         & DT            & HT           \\
        \midrule
        $\mu_\text{n}$    & 2748.461      & 2200.880      & 1376.392     \\
        $\mu_\text{eff}$  & 2748.711      & 2201.140      & 1376.705     \\
        KNM1              & 2748.4600     & 2201.3201     & 1377.0096    \\
        \midrule
                          & $^3$HeT$^{+}$ & $^3$HeD$^{+}$ & $^3$HeH$^{+}$ \\
        \midrule
        $\mu_\text{n}$    & 2748.202      & 2200.714      & 1376.327     \\
        $\mu_\text{eff}$  & 2748.702      & 2201.134      & 1376.702     \\
        KNM1              & 2748.2020     & 2201.1340     & 1376.7023    \\
        \bottomrule
    \end{tabular}
    \label{tbl:reduced_mass:parent_daughter}
\end{table}

\begin{figure}[htpb]
    \centering
    \includegraphics[width=0.60\textwidth, page = 1]{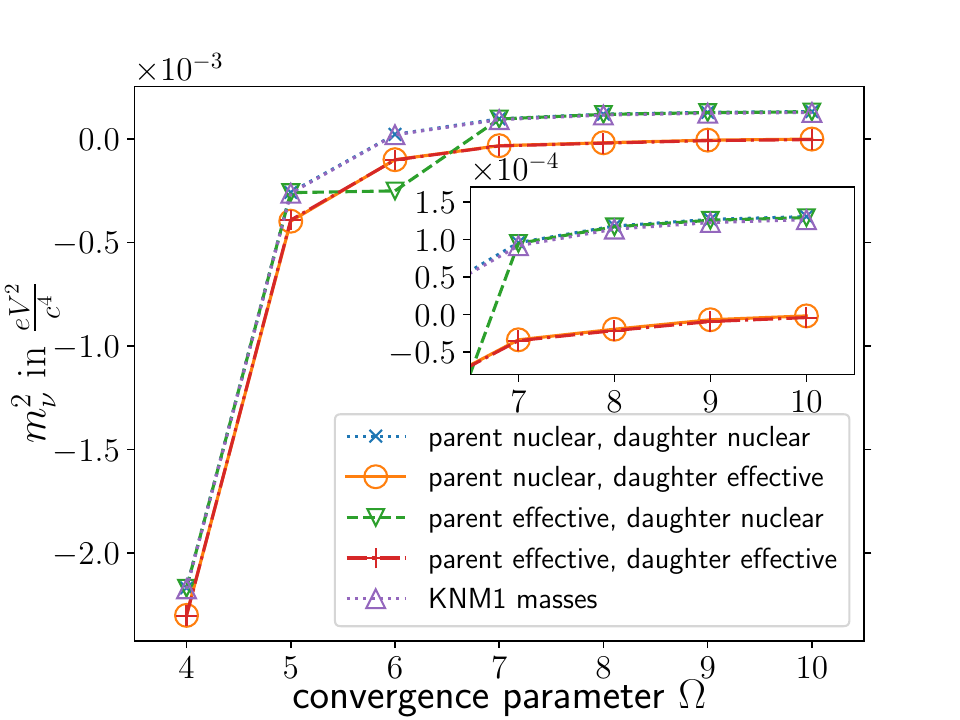}
    \caption{
        Comparison of the fitted squared neutrino mass obtained for different choices of the reduced mass. Shown are the results for different combinations of the adopting the nuclear or the effective reduced mass for either the parent or the daughter molecule, or for both. Furthermore, the results obtained with the reduced masses adopted in the KNM1 FSD calculation are shown.
    }
    \label{fig:results:masses_shift_mass}
\end{figure}

The corresponding values for the reduced masses in the two approximations that were used in the present work are listed in tab.~\ref{tbl:reduced_mass:parent_daughter}. Furthermore, the values used in the KNM1-FSD calculation are also given. 
As it turns out, there appear to be some minute inconsistencies with respect to the adopted reduced masses in the KNM1-FSD calculation.\footnote{For both $\TT$ and $\HeT$ the nuclear reduced masses were evidently adopted. While this is consistent for this isotopologue in itself, it ignores the findings in \cite{dia:doss_thesis}. In fact, this choice is inconsistent with the fact that for $\HeD$ and $\HeH$ the effective reduced masses were evidently adopted. Even more confusing appears the choice of the reduced masses for $\DT$ and $\HT$ in the KNM1-FSD calculation, since they do neither agree to the nuclear nor the effective reduced masses, nor to any of the other reduced masses discussed in \cite{dia:doss_thesis}. Since these values could not be reproduced, it is also possible that some typo occurred in the input data.} However, the effect of the likely erroneous values for $\DT$ and $\HT$ entering the KNM1 FSD should be negligible for two reasons. First, the differences between the values used and the ones that should be used when using the effective reduced masses are very small. Second, the FSDs of $\DT$ and $\HT$ are of much smaller importance compared to the one of $\TT$ because of the high $\TT$ purity of \SI{97.6}{\percent} in the first KATRIN science campaign \cite{KNM1Paper}. Evidently, for the same reason (high $\TT$ purity) the use of the nuclear instead of the effective reduced mass for $\HeT$ appears more \AS{relevant}. The difference between $\mu_\text{n}$ and $\mu_\text{eff}$ is, however, relatively small.

In order to estimate the uncertainty  on $m_{\nu}^2$ due to the choice of the reduced mass, convergence curves for five different mass combinations are considered: the original masses as used in the KNM1 FSD and all four combinations of $\mu_\text{n}$ or $\mu_\text{eff}$ for the parent and daughter isotopologues. This should provide a conservative upper limit on the effect than an inaccurate choice of $\mu$ has on the FSD. In fig.~\ref{fig:results:masses_shift_mass} these five convergence studies are shown.

The choice of the reduced mass for the parent molecule has a negligible effect onto $m_\nu^2$ that is of the order of less than $\SI{1e-5}{\square\electronvolt \per c^4}$ and thus not resolved on the scale used in fig.~\ref{fig:results:masses_shift_mass}. This is to be expected, since the electrons are distributed equally between both nuclei in the electronic ground state of $\TT$. Within the Born-Oppenheimer approximation this applies also to $\DT$ and $\HT$. Nevertheless, adopting the (smaller) nuclear reduced mass compared to the effective reduced mass lowers in first order all energy levels by a constant value.  
The choice of the reduced mass for the daughter isotopologues has a larger effect since it accounts for the unequal distribution of the electron density around the two nuclei. The shift between the $m_\nu^2$ obtained using the nuclear reduced mass and the one obtained using the effective reduced mass for the daughter molecule yields (independent of the choice of the reduced mass of the parent molecule) at $\Omega=10$ an uncertainty estimate of $\SI{1.3e-4}{\square\electronvolt\per c^4}$ due to the choice of reduced masses. The same uncertainty is found when using the reduced masses adopted in the calculation of the KNM1 FSD.   

%--------------------------------------------------------------------------------
% endpoint reference
%--------------------------------------------------------------------------------
\subsubsection{Molecular ground-state energies \label{sec:endpoint-reference}}
Since the molecular tritium source in KATRIN contains impurities like the tritium containing isotopologues $\DT$ and $\HT$, the FSD used in the fit needs to be an incoherent superposition of the FSDs for the different isotopologues, weighted with the composition probability (see sec. \ref{sec:betaspectrum}). When \AS{combining} these different FSDs it is important to properly adjust the three energy scales, {\it i.\,e. }to use a consistent energy zero. \AS{If the parent molecule is initially in its absolute ground state, the energy available to the $\upbeta$ electron is maximised when the daughter molecular ion is also left in its absolute ground state and the neutrino is at rest.} The corresponding molecular ground-state energies are different for the different isotopologues even within the Born-Oppenheimer approximation, since the zero-point energy depends on the reduced mass. As a consequence, the maximum energy available to the $\upbeta$ electron is isotopologue dependent. The corresponding relative energy difference needs to be accounted for when creating the incoherent superposition of the FSDs for the three isotopologues.

\begin{figure}[htpb]
    \centering
    \includegraphics[width=0.60\textwidth, page = 8]{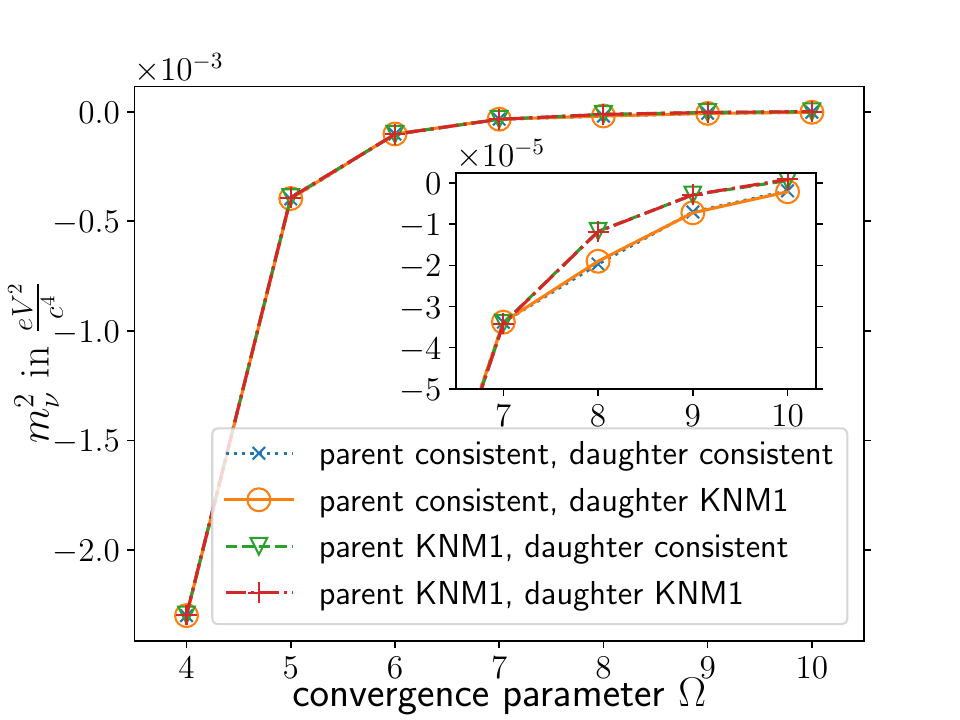}
    \caption{
      Comparison of the extracted squared neutrino mass $m_\nu^2$ for different choices of the ground-state energies of the parent and the daughter molecules in  the adjustment of the energy scales for the different isotopologues. The results obtained with the potential curves used on the KNM1-FSD calculation are independent of $\Omega$ and denoted by "KNM1". If the $\Omega$-dependent ground-state energies from the calculation of the test FSDs are used in the scale adjustment, the results denoted "consistent" are used.       
    }
    \label{fig:results:zero_point}
\end{figure} 
 
Since the value obtained for the absolute ground-state energy depends on the adopted potential curve, uncertainties in the calculated potential curves need to be considered when adjusting the energy scales of the different isotopologues relative to each other. 
In order to determine the impact of uncertainties in the energy-scale adjustment on $m_{\nu}^2$, either the ground-state potential curves used in the KNM1 FSD or the ones obtained within the $\Omega$ variation are used when  adding the FSDs of the different isotopologues in the generation of the test FSDs.
The effect is investigated for daughter and parent ground-state energies separately. Thus four combinations of ground-state energies are obtained, for which the convergence curves are shown in fig.~\ref{fig:results:zero_point}. 
Since the choice of the ground-state energies in the energy-scale adjustment does not cause a constant shift in $m_\nu^2$ as a function of $\Omega$, the uncertainty is conservatively estimated with the maximal difference between the different options considered here which is found for $\Omega = 8$. The uncertainty estimated this way is smaller than $\SI{1e-5}{\square\electronvolt\per c^4}$ for both parent and daughter molecules.

%--------------------------------------------------------------------------------
% binning
%--------------------------------------------------------------------------------
\subsubsection{Binning\label{subsec:binning}}
\begin{figure}
    \centering
    \includegraphics[width=0.6\textwidth]{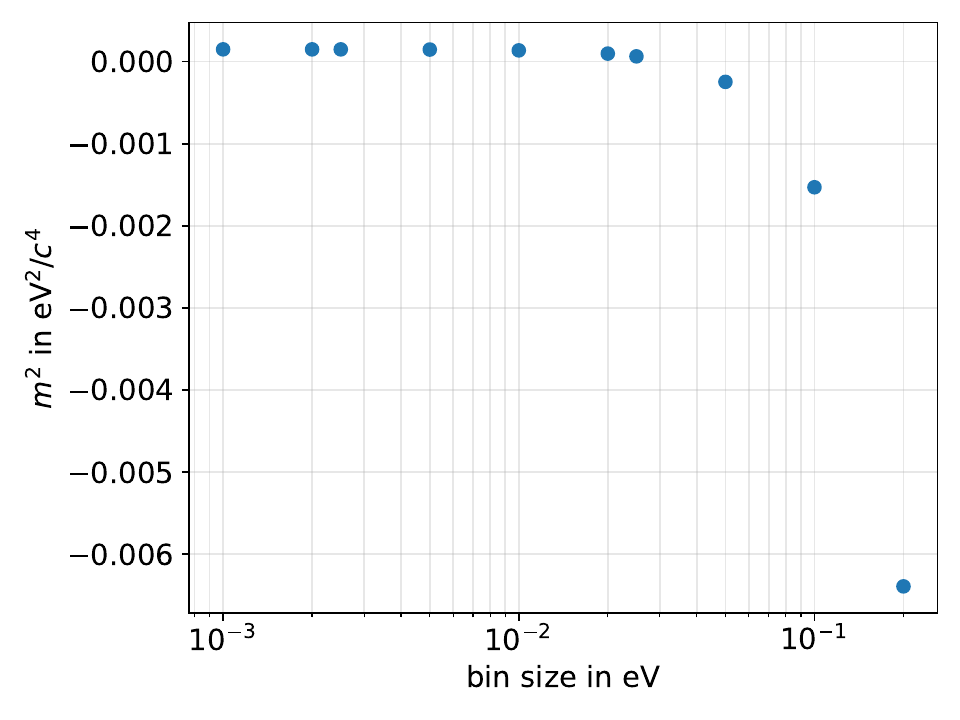}
    \caption{Squared neutrino mass $m_\nu^2$ obtained by the fit as a function of the size of the bins used for binning the FSD.}
    \label{fig:binning}
\end{figure}

The molecular transition probability $P_{fi}^{\rm TS}$ in eq.~\ref{eq:molecular_fsd:molecular_matrix_element} comprises of two parts. First, a set of discrete values describing the transitions to the bound states of $\HeT$ (or its isotoplogues). Second, a continuous transition probability (per energy) describing the transitions into dissociative and (or) ionisation continua. As was explained in sec.~\ref{sec:betaspectrum} such a spectrum is impractical for the analysis of a neutrino-mass experiment like KATRIN. Instead, a binned probability spectrum is used in the fit model when extracting the squared neutrino mass. The influence of equidistant bin sizes on $m_\nu^2$ is investigated in fig. \ref{fig:binning}. The Monte-Carlo data set which serves as reference spectrum uses the pseudo KNM1 FSD with non-equidistant bins. The KNM1 FSD is binned with $\SI{0.02}{\electronvolt}$ wide bins between $\SI{-0.5}{\electronvolt}$ and $\SI{5.0}{\electronvolt}$ (covering the transitions to the electronic ground state of the daughter molecular ion) and $\SI{0.2}{\electronvolt}$ wide bins between $\SI{19}{\electronvolt}$ and $\SI{40}{\electronvolt}$ (electronically excited states). Negative energies can occur due to the non-zero temperature and the combination of different isotopologues since the zero energy is defined via the ground state of T$_2$. For test bin sizes up to \SI{0.025}{\electronvolt}, a small constant positive shift in $m_\nu^2$ around $\SI{1e-4}{\square\electronvolt\per c^4}$ occurs. With increasing bin size, the $m_\nu^2$ shift becomes more negative when the test FSD bin size becomes
larger than the pseudo FSD bin size. This (non-equidistant) binning of the KNM1 FSD is found to yield a shift in $m_\nu^2$ of \SI{1.5e-4}{\square\electronvolt\per c^4} at $\Omega=10$. In conclusion, the binning of a future FSD should not exceed bin sizes of \SI{0.02}{\electronvolt} to keep the uncertainty caused by binning of the FSD on the $10^{-4}$ level.

%--------------------------------------------------------------------------------
% endpoint uncertainty
%--------------------------------------------------------------------------------
\subsubsection{Tritium $Q$ value\label{endpointuncertainty}}
The maximum fractional recoil $\tilde{K}$ (see secs.~\ref{sec:transition_probabilities} and \ref{sec:deliberate_approximations}) imparted by the $\upbeta$ electron and the neutrino onto the internal degrees of freedom of the molecule occurs, if the electron has maximum kinetic energy, {\it i.\,e. }at the end point $E_0$ of the spectrum. This value depends on the energy release in the $\upbeta$ decay and thus on the tritium $Q$ value \jannis{\cite{dia:bodine2015}}. The uncertainty on the $Q$ value for tritium $\upbeta$ decay is dominated by the mass difference between T and $^\text{3}$He. As mentioned in sec.~\ref{sec:conversion_factors}, the KNM1 FSD adopted the value $E_0=$\SI{18573.24}{\electronvolt} for the end-point energy from \cite{dia:bodine2015}, and a corresponding mass difference between T and $^\text{3}$He of \SI{18591.3(1)}{\electronvolt}. A more recent experimental value for the mass difference stems from Penning-trap measurements, by which it is determined to be \SI{18592.01\pm0.07}{\electronvolt} \cite{PhysRevLett.114.013003}. The $Q$ value specific for KATRIN which also includes plasma effects and work function differences with higher uncertainty \cite{KNM1Paper} is not used in the FSD calculation, because these effects have no impact on the FSD uncertainty as they only influence the $\upbeta$-decay electron which passes the whole beamline including the spectrometer. 

\begin{figure}
    \centering
    \includegraphics[width=0.65\textwidth]{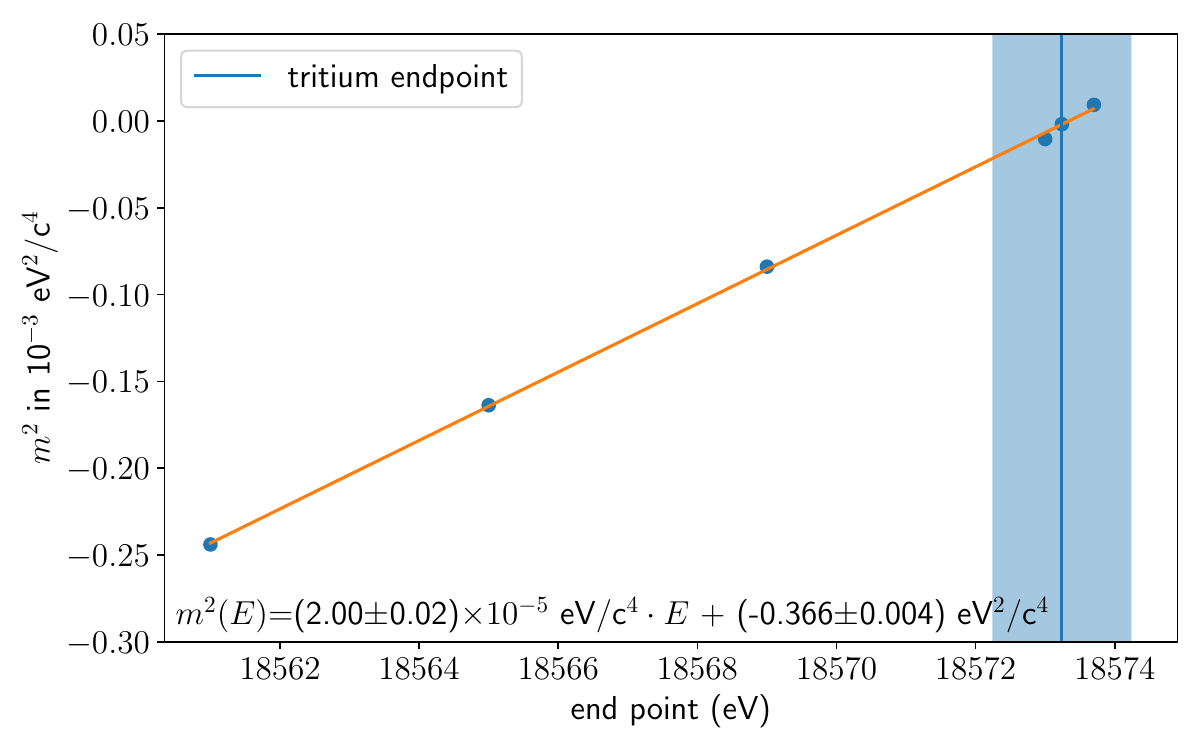}
    \caption{Influence of the choice of the tritium $Q$ value on $m_\nu^2$. The tritium end point used in the FSD calculation is marked as vertical line, together with an uncertainty band marking the assumed end-point uncertainty of \SI{1}{\electronvolt}. The obtained slope for the change of $m_\nu^2$ as a function of the end-point energy is \SI{2e-5}{\square\electronvolt\per c^4\per \electronvolt}, obtained from a linear fit which is shown in orange}.
    \label{fig:endpoint}
\end{figure}

In order to investigate the impact of an uncertainty of the tritium $Q$ value onto the fractional recoil and thus the calculated FSD, test FSDs with different end-point values (and thus with different fractional recoils) were generated, see also following sec.~\ref{constantfractionalrecoil}). (The influence on the reduced masses was neglected, since it is orders of magnitude smaller.) The experimental end point $E_\text{0}$ of the Monte-Carlo data is kept constant at the value given in sec. \ref{generation}. The influence on the extracted value of $m_\nu^2$ is shown in fig.~\ref{fig:endpoint}. Due to the discrepancy of the mass difference chosen for the FSD and the result from Penning-trap measurements, an uncertainty on the $Q$ value of \SI{1}{\electronvolt} is assumed. This uncertainty induces an uncertainty on $m_\nu^2$ of \SI{2e-5}{\square\electronvolt\per c^4}.

\subsubsection{Fractional-recoil variation over the fit interval}\label{constantfractionalrecoil}
The effect of the $Q$-value uncertainty is similar to the effect of approximating the fractional recoil to be constant over the whole fit interval (see sec.~\ref{sec:deliberate_approximations}). Another extreme assumption would be the use of the fractional recoil corresponding to the end of the fit interval, {i.\,e.\ }40 eV below the end point $E_0$ in the present case. A more reasonable estimate within the constant-fractional-recoil approximation is obtained by adopting the fractional recoil that arises at the mean excitation energy (within the fit interval). Since the fractional recoil depends only on the end-point energy of the isotopologue in question, the uncertainty can be determined with the help of the slope from fig.~\ref{fig:endpoint}. Using the mean excitation energy for the $\TT$ pseudo KNM1-FSD ($\SI{10.74}{\electronvolt}$) this yields an estimated uncertainty on $m_\nu^2$ due to the constant-fractional-recoil approximation of \SI{2.2e-4}{\square \electronvolt \per c^4} for $\Omega=10$.\footnote{In the KNM1-FSD calculation also possible electronic excitations due to the fractional recoil were neglected, see sec.~\ref{sec:deliberate_approximations}).}

%--------------------------------------------------------------------------------
% sudden approximation
%--------------------------------------------------------------------------------
\subsubsection{Corrections to the sudden approximation}
\label{subsec:suddenapprox}
\begin{figure}
    \centering
    \includegraphics[width=0.60\linewidth, page = 12]{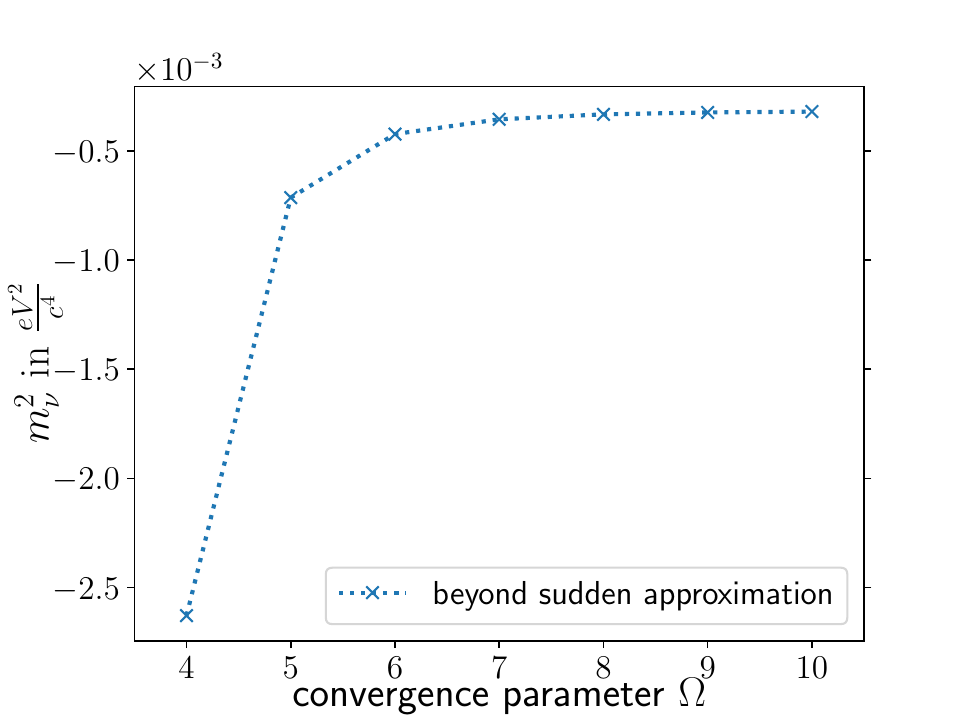}
    \caption{Convergence study for estimating the effects of the corrections to the sudden approximation onto $m_\nu^2$.}
    \label{fig:results:beyond_sudden}
\end{figure}

The first-order correction to the sudden approximation was not only derived and discussed in \cite{PhysRevC56.2162}, but even explicitly evaluated for $\TT$. For a rough estimate of the uncertainty due to the use of the sudden approximation, the total transition probabilities including the first-order correction for the electronic ground and 5 lowest lying excited states given in Table II (column 5) in \cite{PhysRevC56.2162} were used for generating test FSDs.
In more detail, based on the literature values the relative changes of the transition probabilities due to the inclusion of the first-order correction are obtained for the 6 lowest states of $\HeT$. The contributions of these states to the test FSDs that were generated within the sudden approximation are then correspondingly re-scaled with the respective correction factors, yielding new test FSDs. The obtained $m_\nu^2$ extracted from the fits are shown in fig.~\ref{fig:results:beyond_sudden}. The uncertainty due to the use of the sudden approximation is found to be $<\SI{4e-4}{\square\electronvolt\per c^4}$ at $\Omega = 10$.

%--------------------------------------------------------------------------------
% total uncertainty
%--------------------------------------------------------------------------------
\subsection{New FSD uncertainty for the first KATRIN science campaign\label{sec:final-uncertainty}}
Tab. \ref{tab:final_systematics_summary}
summarises all systematic uncertainties on the fitted neutrino mass square $m^2_\nu$ from the calculated FSD used for the KNM1 analysis derived in the studies presented in sec. \ref{sec:individual-contributions}. The first three contributions, ``Basis-set $\Omega$ convergence'', ``Bases for electronically excited states'' and ``Ground-state energies'', are not independent of each other, because they are connected via the variational principle. The corresponding uncertainties are thus added linearly in order to obtain a conservative estimate. Since the other contributions are independent, they are added quadratically, yielding $\Delta m^2_\nu = 5.3 \cdot 10^{-4}$\,eV$^2/c^4$ for the FSD uncertainties. Finally, the assumed uncertainties of the theoretical corrections (adiabatic, relativistic, radiative) are also added quadratically, yielding the total uncertainty of $\Delta m^2_\nu = 1.3 \cdot 10^{-3}$\,eV$^2/c^4$ given in the last row of tab. \ref{tab:final_systematics_summary}.

It is worth mentioning that the temperature fluctuations of $\pm 1$\,K of the KATRIN experiment lead to an additional uncertainty to $m_\nu^2$, which is, however, not related to the FSD calculation itself. A similar argument holds for the gas composition and the ortho-para ratio (only in the case of $\TT$) in the tritium source. An estimate of these experimental uncertainties can be found in app. \ref{app:temperature}. Further deficiencies of the present uncertainty analysis that are due to the use of an imperfect pseudo-KNM1 FSD, namely the inaccuracy due to the discretisation of the dissociation continuum and the incomplete reconstruction of the electronic overlaps, are discussed in app. \ref{app:FSD_parameters}.

\begin{table}
\centering
        \caption{Systematic contributions to the FSD-caused uncertainty of $m_\nu^2$ for the measurement conditions of the first KATRIN science campaign (see text for details on how the total uncertainty is obtained.) The KNM1 bins listed in this table are \SI{2e-2}{\electronvolt} wide bins in the range \SIrange{-0.5}{5}{\electronvolt} and \SI{2e-1}{\electronvolt} wide bins in the range \SIrange{19}{40}{\electronvolt}. \label{tab:final_systematics_summary}}
    \begin{tabular}{p{6.5cm}p{5cm}rp{4cm}}
    \toprule
    contribution                          & input                                                            & $\Delta m^2_\nu$ (eV$^2/c^4$) \\
    \midrule
    Basis-set $\Omega$ convergence        & test of KNM1 FSD vs. FSD with $\Omega=10$                        & $\SI{2e-4}{}$\\
    Bases for electronically excited states & comparison of 4 bases                                            & $\SI{3e-5}{}$\\

    Reduced mass daughter                 & $2748.702\,m_e$ vs. $2748.202\,m_e$ for $^3$HeT$^+$              & $\SI{1.5e-4}{}$\\
    Reduced mass parent                   & $2748.711\,m_e$ vs. $2748.461\,m_e$ for T$_2$                    & $\SI{1e-5}{}$\\
      Ground-state energies                    & comparison of KNM1 v.\ $\Omega$-dependent values                              & $\SI{1e-5}{}$\\
  Binning                               & KNM1 bins vs.\ \SI{0.001}{\electronvolt} bins & $\SI{2e-4}{}$\\
      Tritium Q value           & mass-difference uncertainty of \jannis{$\pm\SI{1}{\electronvolt}$}                    & \SI{2e-5}{}\\
  
    Fractional-recoil variation & $\SI{10.74}{\electronvolt}$ mean-energy to end-point difference & \SI{2e-4}{}\\
        Corrections to sudden approximation   & full inclusion (\SI{100}{\percent})                                               & $\SI{4e-4}{}$\\
    \midrule
    Sum of all contributions by the FSD calculation &                                                        & \SI{5.3e-4}{}\\
    \midrule
    Theoretical corrections (adiabatic, relativistic, radiative)  & partial inclusion (\SI{30}{\percent})                         & $\SI{1.2e-3}{}$\\
    \midrule \midrule
    \textbf{ total uncertainty} &                                                                            & $\SI{1.3e-3}{}$\\
    \bottomrule
    \end{tabular}
\end{table}

%--------------------------------------------------------------------------------
% summary and outlook
%--------------------------------------------------------------------------------
\section{Summary and Outlook}\label{sec:summary}
Motivated by the increased sensitivity of the upcoming measurement campaigns of the KATRIN experiment, the uncertainty of the molecular final-states distribution (FSD) entering the fit of the neutrino mass square $m_\nu^2$ -- up to now conservatively estimated by a fully phenomenological approach -- has been revisited. Studying the impact of the individual sources of uncertainties on the extracted value of $m_\nu^2$ is proposed as a direct and controllable way of obtaining an uncertainty budget that is adapted to the experimental conditions. At the same time this method provides a clear understanding which approximations in the FSD calculations are the limiting ones. A second ingredient of the new scheme is a systematic increase of the basis-set size allowing for a straightforward convergence study. 

The new uncertainty analysis was applied to the FSD used in the analyses of the first two measurement campaigns of the KATRIN experiment, the ``KNM1 FSD''.  The experimental conditions of the first KATRIN campaign were considered in this study. A pseudo-KNM1 FSD being nearly identical to the KNM1 FSD was generated in order to apply the new method including the convergence study based on the size of the molecular basis-set.
The study presented here provides important and detailed insight into both the absolute size and the relative importance of the various factors and approximations entering the FSD evaluation. Most importantly, it is confirmed that the KNM1 FSD is already very accurate with its uncertainty of $\Delta m^2_\nu = 1.3 \cdot 10^{-3}$\,eV$^2/c^4$ derived by this study when applied to the first KATRIN campaign. It was confirmed to be sufficiently accurate for the analyses of the experimental data obtained in the first two neutrino-mass measurement campaigns of KATRIN.            

Based on this work and the pseudo-KNM1 FSD, an FSD for the upcoming KATRIN campaigns is currently under development. This FSD will enable a straightforward calculation of its systematic uncertainties following the methods presented here. The new FSD will also be extended in order to allow analyses of the tritium $\upbeta$-decay spectrum for larger energy intervals, {\it e.\,g.\ }\SI{60}{\electronvolt} below the end point, in the future.

%--------------------------------------------------------------------------------
% appendix
%--------------------------------------------------------------------------------
\begin{appendices}

%------------------------------------------------------------------
% glossary
%------------------------------------------------------------------
\section{Glossary\label{glossary}}

\begin{description}
  \item[KNM1 FSD] was used for the evaluation of the first and second measurement campaigns of KATRIN. It is based  on literature data and its generation is described in \cite{valerian_paper}. 

  \item[pseudo-KNM1 FSD] is an FSD that is produced for the uncertainty study of the KNM1 FSD, because the here proposed uncertainty analysis cannot be performed with the KNM1 FSD itself. The pseudo-KNM1 FSD contains intentionally some further (though small) modifications compared to the KNM1 FSD, as is described in sec. \ref{sec:5.3}.

  \item[test FSD] is a modification of the pseudo-KNM1 FSD describing a specific uncertainty (see eq.\,\eqref{eq:introduction:beta_decay_fermi}). This uncertainty can be a specific limitation of the calculation (like the finite basis set), an adopted approximation (like the sudden approximation), or some input error. 

  \item[Fackler FSD] was produced by Fackler et al.\,\cite{Fackler1985}.

  \item[Saenz FSD] was produced by Saenz et 
  al.\,\cite{PhysRevLett.84.242}. The main differences to the earlier produced Fackler FSD were the use of the relativistic recoil and the way the electronic continuum was treated.

  \item[basis set] is defined by all parameters specifying the basis functions used to expand the electronic wavefunction, i.\,e.\ it is a combination of a base (defined by four non-integer numbers) and a list of quintuples, a quintuple being a set of five integers. Instead of a list of quintuples alternatively a single parameter, the rank of the basis set $\Omega$, is used. In the latter case all quintuples that fulfill the condition that their sum is equal or smaller than $\Omega$ are included in the basis set. 

  \item[basis] is used synonymously for basis set.

  \item[base parameters] are the non-integer exponential parameters $\alpha$, $\bar{\alpha}$, $\beta$ and $\bar{\beta}$ in the Kolos-Wolniewicz basis functions, see eq.\,\eqref{eq:molecular_fsd:kolos_basis_functions}
  
  \item[base] is a complete set of all four base parameters $\{\alpha, \bar{\alpha}, \beta, \bar{\beta}\}$. 

\item[energy scales] Note, there is a confusing issue related to the used energy scale depending on context. When discussing the $\upbeta$-decay spectrum, energies are generally defined in relation to the end point on a negative scale, e.\,g., the lower limit of the fit interval is defined as $E_\text{0}-\SI{40}{\electronvolt}$. In contrary, energies related to the FSD are generally expressed on a positive scale in relation to the chosen zero energy (the transition from the ground state of $\TT$ to the ground state of $\HeT$). Thus, both scales are antiparallel to each other and any directional terms (like above and below) are context dependent.

\end{description}

\section{Parameters entering the FSD calculation\label{app:FSD_parameters}}

%--------------------------------------------------------------------------------
% non-linear basis parameters
%--------------------------------------------------------------------------------
\subsection{Potential curves and electronic transition matrix elements}\label{app:basis}

%--------------------------------------------------------------------------------
% \HH ground state paramaters
%--------------------------------------------------------------------------------    
\subsubsection{$\HH$ ground state bases}
\label{sec:h2_ground_state_parameters}
The bases used for computing the Born-Oppenheimer potential curve of $\HH$ up to $R = 4.8 \, a_0$ were taken from \cite{dia:kolos86}. For the bases between $R = 4.8 \,a_0$ and $R = 12.0 \,a_0$ a linear interpolation of the re-optimised bases for larger internuclear separations as in \cite{dia:woln93} was used. Together with the exponents from \cite{dia:kolos86} the correctness of the linear interpolation and the bases were checked by recomputing the electronic energies.
     
For finding the energy eigenvalues the canonical orthonormalisation  (see, {\it e.\,g.\ }\cite{dia:woln93}) was used. As in \cite{dia:woln93} the variable $\epsilon$ is introduced as a cutoff, {\it i.\,e.\ }eigenvectors corresponding to eigenvalues smaller than $\epsilon$ are omitted. The value $\epsilon = 10^{-12}$ reported in \cite{dia:woln93} was, depending on the internuclear separation R, found to be too small or too big. The ordering of the basis functions (quintuples), which is unknown, can change the resulting eigenvalues. Therefore, another strategy was finally chosen: the number of basis functions in the canonical orthonormalisation was reduced (compared to an eigenvalue cutoff with $\epsilon$) until the lowest energy eigenvalue was judged to be "satisfactory", {\it i.\,e.\ }it did not violate the variational principle. This does not necessarily mean the eigenvalues closest to the ones from \cite{dia:woln93} are obtained, but should be rather consistent for the present order of exponents. 

Note, the quintuples (basis functions) given in \cite{dia:kolos86} contain twice the same quintuple ($i = 131$ and $i  = 215$), so one of them was replaced with $\{1,0,2,2,2\}$ as this quintuple was not yet contained. Following \cite{dia:woln93} also the quintuple $\{1,3,1,1,1\}$ was changed into $\{0,3,1,1,1\}$. Of course, it may be expected that the ordering of the quintuples (basis functions) is irrelevant. However, this is not the case here due to the finite numerical precision used in the Ko{\l}os code.

For all the distances above $R = 0.8\, a_0$ results agreeing with the values in \cite{dia:kolos86} within a precision of $10^{-9}$ to $10^{-8}$ were achieved. Keeping in mind that the original ordering of the basis functions (quintuples) is unknown, this is rather good. For internuclear separations $R < 0.8 \, a_0$ the results are orders of magnitudes worse, despite the knowledge of the bases. However, they do not contribute notably to the FSD.
     
The literature bases cannot be used in a fully unchanged manner in the convergence study, because of the inability of the H2SOLV code \cite{CPC} to process $|a| < 0.01 \cdot R$ where a is any of the four non-linear parameters inside the base and R is the internuclear separation. This leads to the problem that in between $R = 3.4 \, a_0$ and $R = 4.2 \, a_0$ the original values of $\bar{\beta}$ cannot be used. Instead, a scaling of $\bar{\beta}$ to the minimum value required by the H2SOLV code is performed. This leads to the bases shown in tab.~\ref{tbl:molecular_fsd:scaled_h2_parameters}.

\begin{table}[htpb]
    \centering
    \caption{
        Original (left hand side of the table) and modified bases (right hand side of the table) with re-scaled $\bar{\beta}$ to be compatible with the H2SOLV code. The re-scaling was performed according to $\bar{\beta} = 0.01 \cdot R$.
    }
    \begin{tabular}{l|llll|llll}
        \toprule
        R & $\alpha$ & $\bar{\alpha}$ & $\beta$ & $\bar{\beta}$ & $\alpha$ & $\bar{\alpha}$ & $\beta$ & $\bar{\beta}$ \\
        \midrule                  
        3.4 & 2.364 & 2.052 & 1.385 &  0.032 & 2.364 & 2.052 & 1.385 &  0.034 \\
        3.6 & 2.470 & 2.149 & 1.451 &  0.020 & 2.470 & 2.149 & 1.451 &  0.036 \\
        3.8 & 2.568 & 2.253 & 1.521 &  0.006 & 2.568 & 2.253 & 1.521 &  0.038 \\
        4.0 & 2.676 & 2.381 & 1.595 & -0.011 & 2.676 & 2.381 & 1.595 & -0.040 \\
        4.2 & 2.782 & 2.527 & 1.662 & -0.035 & 2.782 & 2.527 & 1.662 & -0.042 \\
        \bottomrule
    \end{tabular}
    \label{tbl:molecular_fsd:scaled_h2_parameters}
\end{table}

%--------------------------------------------------------------------------------
% \HeH ground state parameters
%--------------------------------------------------------------------------------    
\subsubsection{$\HeH$ ground-state bases}\label{sec:heh_ground_state}
The Born-Oppenheimer potential curve for the ground state of $\HeH$ was taken from \cite{dia:kolos85}, \cite{dia:kolo76b}, and \cite{dia:kolo76a}. Additionally, an improvement of the energies of the potential curve is added in form of an interpolated shift, as proposed in \cite{dia:bish79}. Unfortunately, only \cite{dia:bish79} provides the bases parameters, and they are only given for a few selected internuclear separations $R \in \{ 0.9, 1.46, 1.8, 3.0, 4.5, 6.0\}$. Of course, there is no guarantee that any of the other three publications actually used these bases. 

The six bases from \cite{dia:bish79} were interpolated with a polynomial of degree five, thus adopting the same procedure used in \cite{dia:bish79} for generating bases for $R = 1.2 \,a_0$ and $R = 2.4 \, a_0$. 

A way to determine the quintuples defining the basis functions is also given in \cite{dia:bish79}. 
These quintuples were checked at the internuclear separations where the bases are given. The achieved agreement with the literature values is $10^{-8}$ or even $10^{-9}$, that is to all given digits. An exception is the energy at $R = 6.0 \, a_0$ where only a precision of $10^{-7}$ is achieved. Still the quintuples should at least be correct, but may not necessarily be in the correct order.
With these quintuples the interpolated bases for the distances between $R = 0.9 \,a_0$ and $R = 6.0\, a_0$ were checked and an agreement of $10^{-6}$ to $10^{-7}$ was reached. It should be noted that the energies to which the comparison is made are not given explicitly, but rather had to be reconstructed from the aforementioned set of four publications (\cite{dia:kolos85}, \cite{dia:kolo76b}, \cite{dia:kolo76a}, and \cite{dia:bish79}).

The most severe problem is the missing base for $R = 8 \, a_0$ and thus the extension of the bases beyond $R = 6 \,a_0$. It was tried to use the extension of the interpolating polynomial of degree 5 from before, but this failed, especially for larger internuclear distances $R$. The results from two different linear interpolations were compared, once using only the bases at $R = 4.5 \, a_0$ and $R = 6 \, a_0$ and once using all given bases. Furthermore, a simple scaling of the bases at $R = 6 \, a_0$ with $R$ was tested. All three methods give comparable results and are closer to each other than to the original values. The linear interpolation with only two points ($R = 4.5 a_0$ and $R = 6 a_0$) was finally chosen to extrapolate to larger distances.
The bases finally constructed from the polynomial interpolation and the linear extrapolation are reported in tab.~\ref{tbl:molecular_fsd:parameters_heh_ground_state}.
     
\begin{table}[htpb]
    \centering
    \caption{
         Table with the reconstructed KNM1 bases for the electronic ground state of the $\HeH$ molecule. The internuclear separations $R$ are given in units of Bohr ($a_0$).
     }
    \begin{tabular}{l|llll|l|llll}
        \toprule
        R & $\alpha$ & $\bar{\alpha}$ & $\beta$ & $\bar{\beta}$ & R & $\alpha$ & $\bar{\alpha}$ & $\beta$ &  $\bar{\beta}$ \\
        \midrule
        0.6  & 0.55   & 0.7716 & 0.2104 & 0.02363 & 2.8 & 2.752 & 2.495 & 2.846 & 1.458 \\
        0.8  & 0.8698 & 0.8633 & 0.4214 & 0.184   & 3.0 & 2.973 & 2.64  & 3.095 & 1.497 \\
        0.9  & 1.004  & 0.926  & 0.53   & 0.281   & 3.2 & 3.205 & 2.786 & 3.341 & 1.553 \\
        1.0  & 1.125  & 0.9971 & 0.6406 & 0.384   & 3.4 & 3.446 & 2.936 & 3.582 & 1.637 \\
        1.1  & 1.234  & 1.075  & 0.7531 & 0.4897  & 3.5 & 3.568 & 3.014 & 3.701 & 1.693 \\
        1.2  & 1.334  & 1.158  & 0.8675 & 0.5953  & 3.6 & 3.692 & 3.094 & 3.819 & 1.761 \\
        1.3  & 1.427  & 1.244  & 0.9837 & 0.6985  & 3.8 & 3.94  & 3.263 & 4.05  & 1.932 \\
        1.4  & 1.515  & 1.333  & 1.102  & 0.7974  & 4.0 & 4.188 & 3.446 & 4.276 & 2.156 \\
        1.46 & 1.565  & 1.387  & 1.173  & 0.854   & 4.5 & 4.789 & 3.969 & 4.821 & 2.954 \\
        1.5  & 1.598  & 1.423  & 1.221  & 0.8904  & 5.0 & 5.349 & 4.569 & 5.354 & 3.986 \\
        1.6  & 1.679  & 1.514  & 1.342  & 0.9765  & 5.5 & 5.898 & 5.153 & 5.913 & 4.903 \\
        1.7  & 1.759  & 1.604  & 1.464  & 1.055   & 6.0 & 6.537 & 5.524 & 6.56  & 5.026 \\
        1.8  & 1.838  & 1.694  & 1.587  & 1.125   & 6.5 & 7.12  & 6.042 & 7.14  & 5.717 \\
        1.9  & 1.918  & 1.782  & 1.711  & 1.187   & 7.0 & 7.702 & 6.561 & 7.719 & 6.407 \\
        2.0  & 1.999  & 1.869  & 1.836  & 1.24    & 7.5 & 8.285 & 7.079 & 8.299 & 7.098 \\
        2.2  & 2.169  & 2.036  & 2.088  & 1.324   & 8.0 & 8.868 & 7.597 & 8.879 & 7.789 \\
        2.4  & 2.35   & 2.196  & 2.341  & 1.383   & 9.0 & 10.03 & 8.634 & 10.04 & 9.17  \\
        2.6  & 2.544  & 2.348  & 2.594  & 1.424   & 10  & 11.2  & 9.671 & 11.2  & 10.55 \\
        \bottomrule
    \end{tabular}
\label{tbl:molecular_fsd:parameters_heh_ground_state}
\end{table}
     
%--------------------------------------------------------------------------------
% \HeH excited state parameters
%--------------------------------------------------------------------------------    
\subsubsection{$\HeH$ excited-state bases\label{basis}}
For the excited states of $\HeH$ the potential curves were taken from \cite{PhysRevC.60.034601}. These constitute a compromise of the energies and overlaps obtained with four different bases. Two of the bases are given in \cite{dia:saenz93} and in accordance to later publications the naming of Base 1 and Base 2 was swapped respectively (named Basis 1 and Basis 2 in \cite{dia:saenz93}). The third base (Base 3) is not known and was therefore reconstructed. The potential curves from \cite{PhysRevC.60.034601} were used together with the knowledge of Base 1 and Base 2 to determine the points that had been computed with Base 3. With these points it was possible to reconstruct Base 3 as $\{\alpha = 2.08543$, $\bar{\alpha} = 0.68235$, $\beta = 0.79449$, $\bar{\beta} = 0.54546\}$ which gives results within a precision of $10^{-7}$ to $10^{-8}$ compared to \cite{PhysRevC.60.034601}.  The energies and overlaps for $R = 3 \, a_0$ and $R = 4 \, a_0$ in \cite{PhysRevC.60.034601} were taken from \cite{dia:kolos85}, where, however, no bases or quintuples for their computation are given. This means the fourth base is not known and cannot be reconstructed in a reasonable time. Instead, the parameters for the ground state of $\HeH$ (see app.~\ref{sec:heh_ground_state}) were taken at $R = 3 a_0$ and $R = 4 a_0$, as these should give at least comparable results for the excited states.

%--------------------------------------------------------------------------------
% electronic overlaps
%--------------------------------------------------------------------------------
\subsubsection{Electronic overlaps\label{overlap}}
For the computation of the electronic overlaps the bases for the ground state of $\HH$ and all states of $\HeH$ (contributing in the fit interval) are needed. As the electronic overlaps used in the KNM1 FSD are taken from \cite{dia:kolos85} (which predates \cite{dia:kolos86} from which the $\HH$ bases were taken), it cannot be assumed that in both cases the same bases were used. For $\HeH$ it is assumed that the same bases have been used, because they were taken from \cite{dia:bish79}, which acknowledges Ko{\l}os for these bases. For the $\HH$ bases the values from \cite{dia:kolo64} up to and including $R = 2.0 \,a_0$ were taken. For larger distances the bases used in \cite{dia:prc99} \footnote{The numerical data were provided by A.~Saenz.} were taken, which are listed in tab.~\ref{tbl:molecular_fsd:overlap_parameters_h2}.
For the electronic overlaps with the $\HH$ ground state, only the bases at $R \in \{3.0, 4.0\}\, a_0$ from tab.~\ref{tbl:molecular_fsd:overlap_parameters_h2} were used. For the other internuclear separations $R \in \{0.6, 0.8, 1.0, 1.2, 1.4, 1.6, 1.8, 2.0\}\, a_0$ the bases were taken from \cite{dia:kolo64}.

\begin{table}
    \caption{
        Additional $\HH$ bases used in the electronic overlap calculation, complementing the ones given in \cite{dia:kolo64}.
    }
    \centering
    \begin{tabular}{l|llll}
        \toprule
        R    & $\alpha$ & $\bar{\alpha}$ & $\beta$ & $\bar{\beta}$ \\
        \midrule
        2.2  & 1.406    & 1.406          & 0.263   & 0.263 \\
        2.4  & 1.499    & 1.499          & 0.328   & 0.328 \\
        2.6  & 1.575    & 1.575          & 0.375   & 0.375 \\
        2.8  & 1.680    & 1.680          & 0.450   & 0.450 \\
        3.0  & 1.780    & 1.780          & 0.519   & 0.519 \\
        4.0  & 2.300    & 2.300          & 0.848   & 0.848 \\
        \bottomrule
    \end{tabular}
    \label{tbl:molecular_fsd:overlap_parameters_h2}
\end{table}

\begin{figure}
    \centering
    \includegraphics[width=0.60\textwidth, page = 11]{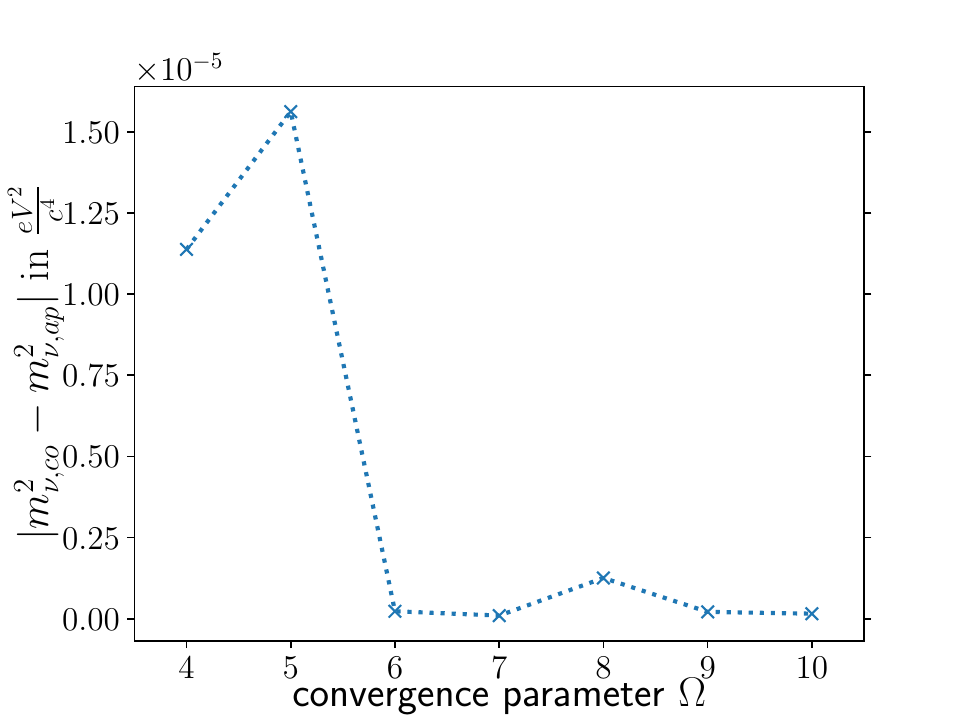}
    \caption{
        Absolute difference between the $m_\nu^2$ fit results obtained with the overlaps used in the convergence studies and the more appropriately chosen overlaps. Here $m^2_{\nu,co}$ stands for the shift obtained with the overlaps from the convergence study and $m^2_{\nu,ap}$ is the shift obtained with more appropriately chosen $\HH$ bases.
    }
    \label{fig:results:correct_overlaps}
\end{figure}

Since different bases for $\HH$ were chosen in the computation of the electronic overlaps than for the computation of the potential curves, an inconsistency is introduced. This affects the ground states and the excited states and thus, in principle, the computation of the complete FSD. As the electronic overlaps for the ground states use an even lower number of internuclear separations $R$ than for the excited states and the ground state makes up roughly 57\% of the total probability, the effect of using the $\HH$ bases from \cite{dia:woln93} was investigated.
These should be more accurate, as they use more points (not all points were used, but rather a selection of 24 points between $R = 0.6$ and $R = 10.0$).
In fig.~\ref{fig:results:correct_overlaps} the absolute difference between the convergence curves using the overlaps used in the convergence studies and the more appropriately chosen overlaps, {\it i.\,e.\ }where the $\HH$ bases from \cite{dia:woln93} were used (which are also used for the computation of the potential curves), is shown. 

As shown in fig. \ref{fig:results:correct_overlaps}, the difference in the fitted $m_\nu^2$ is at worst about $<\SI{2e-5}{\square\electronvolt\per c^4}$, at $\Omega = 5$. So the effect of using inferior bases for the calculation of the electronic overlaps is negligible compared to the other uncertainties. 

%--------------------------------------------------------------------------------
% dissociation continuum
%--------------------------------------------------------------------------------
\subsection{Description of the dissociation continuum\label{app:dissociation}}
As mentioned in sec. \ref{sec:nuclear-motion}, all solutions of the nuclear-motion problem for the electronically excited states of $\HeT$ describe dissociative and thus continuum states. In the evaluation of the KNM1 FSD this problem was treated with a discretisation approach. Adopting a very large $B$-spline basis and fixed-boundary conditions results in a very large number of discretised continuum states. This is, however, computationally very expensive, since a very large eigenvalue problem needs to be solved. This has to be repeated for all the electronically excited states obtained with a given two-electron basis, in the present case depending on the value of $\Omega$, as discussed above in sec.~\ref{omega-variation}.  

Computationally less expensive is the so-called density approach in which the discretised continuum states are continuum re-normalised as to yield probability densities per energy (at a discrete set of energy values). These densities can be interpolated and binned to obtain the FSD.
In order to save computational resources and time, all the FSDs calculated in this work were obtained with the aid of the density approach, since this reduces the computation time by roughly a factor of 6. Clearly, it needs to be checked that this does not introduce some non-negligible error. Therefore, a separate study was made in which all other parameters are chosen as they were used in the KNM1 FSD calculation, but performing the calculation once with the original discretisation approach and once with the density approach. Noteworthy, this is a check of the calculation of the test and the pseudo-KNM1 FSDs obtained in this work and whether uncertainties found when comparing to the KNM1 FSD stem from the different way of treating the dissociation continuum of the electronically excited states. Therefore, the result does not influence the results obtained in the analysis of the first science campaign of KATRIN adopting the KNM1 FSD calculated with the more time-consuming discretisation approach and thus does not enter the uncertainty budget, but is only a check whether the conclusions of the present work are biased by using a different (computationally less expensive) way of treating the dissociative continuum of the electronically excited states.  
The impact of using the discretised approach or the density approach is evaluated in the following way: 
An additional pseudo-KNM1 FSD is calculated, where the discrete instead of the density approach is used. The Monte Carlo data set is fitted with this newly calculated FSD, thus delivering the difference between the two approaches. The found discrepancy is $\SI{1e-7}{\square\electronvolt\per c^4}$, which means this difference is negligible and thus the here adopted density approach does not bias the findings of this uncertainty study.

%--------------------------------------------------------------------------------
% temperature
%--------------------------------------------------------------------------------
\section{Influence of experimental uncertainties entering the FSD calculation}\label{app:temperature}
\begin{figure}
\centering
   \includegraphics[width=0.6\textwidth]{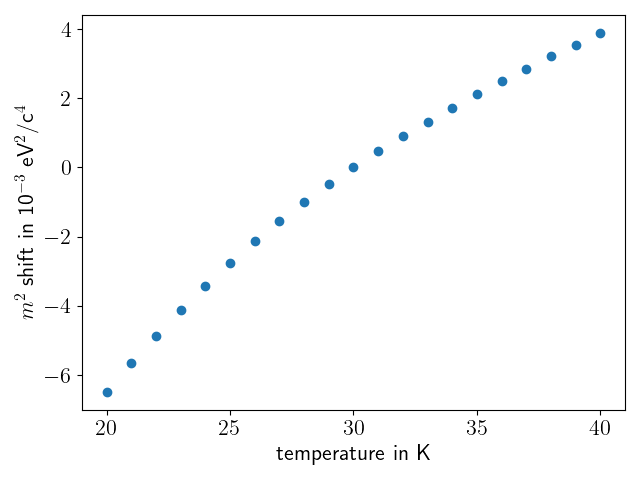}
    \caption{Impact of source temperature on $m_\nu^2$. The slope at $\SI{30}{\kelvin}$ is $\SI{4.8e-4}{\square \electronvolt \per c^4 \per \kelvin}$.\label{temperature}}
\end{figure}

\paragraph{\textit{Temperature}}
The tritium source of KATRIN features a temperature stability of $\pm\SI{1.5}{\milli\kelvin\per\hour}$ at \SI{30}{\kelvin}, corresponding to  $\pm\SI{4.5}{\milli\kelvin}$ during one run of typically \SI{3}{\hour} duration, and a temperature homogeneity of \SI{0.85}{\kelvin} \cite{WGTSTemperature}, leading to a total temperature uncertainty of $<\SI{1}{\kelvin}$. The FSD-related effect of temperature deviations on $m_\nu^2$ is investigated by creating FSDs with angular momenta weighted for different temperatures and using them for a fit on the reference Monte Carlo data with the pseudo-KNM1 FSD generated at \SI{30}{\kelvin}. As demonstrated in fig. \ref{temperature}, the absolute effect on $m_\nu^2$ caused by temperature variations of $\pm \SI{1}{\kelvin}$ is $<\SI{5e-4}{\square\electronvolt\per c^4}$.

\paragraph{\textit{Isotopologue composition}}
The isotopologue composition of the tritium gas within the WGTS is monitored using Raman Spectroscopy \cite{LARA}. The effect of the gas composition on $m_\nu^2$ is probed via the $m_\nu^2$ shift induced by another gas composition used for the fit model applied to the pseudo-KNM1 Monte Carlo data set which is generated with the KNM1 gas composition. Both the T$_2$ purity as well as the HT$/$DT-ratio are tested. Based on \cite{LARA}, an uncertainty of $0.002$ on the tritium purity and $0.04$ on the HT$/$DT-ratio are assumed. The obtained shift in $m_\nu^2$ is $\SI{4e-5}{\square\electronvolt\per c^4}$ due to the tritium purity and $\SI{1e-6}{\square\electronvolt\per c^4}$ due to the HT$/$DT-ratio. It can be concluded that the uncertainty on $m_\nu^2$ originating from the gas composition is negligibly small.

\paragraph{\textit{Ortho-para ratio}}
For the KATRIN tritium source, an ortho-to-para-tritium ratio $\lambda=0.75$, corresponding to a ratio of $3:1$, is assumed, which is the equilibrium ratio at room temperature \cite{AnalysisPaper}. The equilibrium ratio at $\SI{30}{\kelvin}$ is $\lambda=0.57$. However, since the equilibration process is slow relative to the short retention time of $\mathcal{O}(\text{s})$ of the tritium in the WGTS, the $\SI{30}{\kelvin}$ equilibrium is not reached. Estimations based on catalytic conversion times for tritium\footnote{Estimation by P.A. Krochin Yepez, L. Kuckert, M. Kleesiek, K. Valerius, M. Schlösser}\cite{Albers} lead to a worst-case upper limit on the ortho-para conversion in the WGTS of $\SI{3}{\percent}$ and therefore $\lambda=0.7275$.\\ The impact of a $\SI{3}{\percent}$ difference in the ortho-para ratio is investigated by generating a test FSD composed for an ortho-para ratio of 0.7275 and using it for the fit model applied to the pseudo-KNM1 Monte Carlo data set. The Monte Carlo data set assumes an ortho-para ratio of 0.75. The observed shift in $m_\nu^2$ is $\SI{7.1e-4}{\square\electronvolt\per c^4}$ which can be seen as a worst-case shift caused by the ortho-para ratio.

%--------------------------------------------------------------------------------
% comparison KNM1 FSD - pseudo KNM1 FSD
%--------------------------------------------------------------------------------
\section{Comparison of results for $m_\nu^2$ using KNM1 FSD and pseudo-KNM1 FSD\label{sec:fsd-comparison}}
A fit using the KNM1 FSD on the reference MC data set with the pseudo-KNM1 FSD yields a shift in $m_\nu^2$ of $\SI{-3.5e-5}{\square\electronvolt\per c^4}$. Since the calculation of the masses for the KNM1 FSD and the pseudo-KNM1 FSD differs, it would at first glance be expected to see the shift of \SI{1.3e-4}{\square\electronvolt\per c^4} between the two FSDs which is demonstrated in fig. \ref{fig:results:masses_shift_mass}. However, there is an additional shift between the two FSDs which counteracts the shift from the different masses: the shift of \SI{-1.7e-4}{\square\electronvolt\per c^4} from the unconverged potential curves of the KNM1 FSD which is shown in fig. \ref{fig:results:omega_convergence}. The change from the discrete approach for the description of the dissociation continuum in the KNM1 FSD to the density approach for the pseudo-KNM1 FSD causes another shift of \SI{1e-7}{\square\electronvolt\per c^4}. This leaves a discrepancy of \SI{5e-6}{\square\electronvolt\per c^4} between the KNM1 FSD and pseudo-KNM1 FSD, which can be assigned to the imperfect reproducibility of the KNM1 FSD due to missing computational details in the adopted literature values.

\end{appendices}
\newpage
\section*{Declarations}
\paragraph{Funding}
This work was supported by the German Ministry for Education and Research BMBF (05A20PMA) and Deutsche Forschungs\-gemeinschaft DFG (Research Training Group GRK 2149).
\paragraph{Conflict of interest/Competing interests}
The authors have no relevant financial or non-financial interests to disclose.

\paragraph{Acknowledgements}
We kindly thank Andrei Krochin and Magnus Schlösser for their estimations on the ortho-para ratio of molecular tritium in the source, as well as Robin Größle for his inputs related to this topic. We further acknowledge Thierry Lasserre, Diana Parno, Helmut Telle and Larisa Thorne for their valuable comments on the manuscript.

\paragraph{Availability of data and materials}
The datasets generated during and/or analysed during the current study are available from the corresponding author on reasonable request.
\paragraph{Code availability}
Not available.

\bibliography{bibliography.bib}
\end{document}